\documentclass[lettersize,journal]{IEEEtran}
\usepackage{amsmath,amsfonts}
\usepackage{algorithmic}
\usepackage{algorithm}
\usepackage{array}
\usepackage[caption=false,font=normalsize,labelfont=sf,textfont=sf]{subfig}
\usepackage{textcomp}
\usepackage{stfloats}
\usepackage{url}
\usepackage{verbatim}
\usepackage{graphicx}
\usepackage{cite}
\usepackage{array}
\usepackage{amssymb}
\usepackage{textcomp}
\usepackage{stfloats}
\usepackage{url}
\usepackage{verbatim}
\usepackage{graphicx}
\usepackage{cite}
\usepackage{cuted}
\usepackage{bm}
\usepackage{calligra}

\usepackage{amsmath} 
\usepackage{booktabs}
\usepackage{epsfig}
\usepackage{epstopdf}
\usepackage{amsmath}
\usepackage{subfig,graphicx}
\usepackage{multirow}
\usepackage{epsfig}
\usepackage{amsmath}
\usepackage{epstopdf}
\usepackage[justification=centering]{caption}
\usepackage{color}

\begin{document}

\title{Computationally Efficient Unsupervised Deep Learning for Robust Joint AP Clustering and Beamforming Design in Cell-Free Systems}

\author{Guanghui~Chen,~\IEEEmembership{Graduate~Student~Member,~IEEE}, Zheng~Wang,~\IEEEmembership{Senior~Member,~IEEE}, Hongxin~Lin, Yongming~Huang,~\IEEEmembership{Senior~Member,~IEEE}, Luxi~Yang,~\IEEEmembership{Senior~Member,~IEEE}
	% <-this % stops a space
	%\thanks{This work was supported the National Natural Science Foundation of China under Grants 61720106003, 61901107,is U1936201, and 61971128, the Research Project of Jiangsu Province under Grant BE2018121, the Natural Science Foundation of Jiangsu Province under Grant BK20190337, the Natural Science Foundation of Fujian Province under Grant 2019J01055, the National Key R\&D Program of China under 2018YFB1800801 and the Fundamental Research Funds for the Central Universities 2242021R41114. (Corresponding authors: Yongming Huang, Shiwen He).}% <-this % stops a space
	\thanks{This work was supported by the National Natural Science Foundation of China under Grant 62225107, the Natural Science Foundation on Frontier Leading Technology Basic Research Project of Jiangsu under Grant BK20222001, the Fundamental Research Funds for the Central Universities under Grant 2242022k60002. }
	
	\thanks{G. Chen, Z. Wang, Y. Huang and L. Yang are with the School of Information Science and Engineering, and the National Mobile Communications Research Laboratory, Southeast University, Nanjing 210096, China; Y. Huang and L. Yang  are also with the Pervasive Communications Center, Purple Mountain Laboratories, Nanjing 211111, China. (e-mail:cgh@seu.edu.cn, wznuaa@gmail.com, huangym@seu.edu.cn, lxyang@seu.edu.cn).}
	%\thanks{S. He is with the School of Software, Xinjiang University, Urumqi, 830049, Xinjiang, China and is also with the School of Computer Science and Engineering, Central South University, Changsha 410083, China. (email: shiwen.he.hn@csu.edu.cn).}
	%\thanks{S. He is with the School of Computer Science and Engineering, Central South University, Changsha 410083, China. S. He is also with the Pervasive Communications Center, Purple Mountain Laboratories, Nanjing 211111, China. (email: shiwen.he.hn@csu.edu.cn).}
	\thanks{H. Lin is with the Purple Mountain Laboratories, Nanjing 211111, China. (email: linhongxin@pmlabs.com.cn).}
	
	%\thanks{Y. Huang and L. Yang are with the School of Information Science and	Engineering, and the National Mobile Communications Research Laboratory, Southeast University, Nanjing 210096, China are also with the Pervasive Communications Center, Purple Mountain Laboratories, Nanjing 211111, China. (e-mail: huangym@seu.edu.cn, lxyang@seu.edu.cn).}
}



\maketitle

\begin{abstract}
In this paper, we consider robust joint access point (AP) clustering and beamforming design with  imperfect channel state information (CSI) in cell-free systems. Specifically, we jointly optimize AP clustering and beamforming with imperfect CSI to simultaneously maximize the worst-case sum rate and minimize the number of AP clustering under power constraint and the sparsity constraint of AP clustering. By transformations,  the semi-infinite constraints caused by the imperfect CSI are converted into more tractable forms for facilitating a computationally efficient unsupervised deep learning algorithm. In addition, to further reduce the computational complexity, a computationally effective unsupervised deep learning algorithm is proposed  to implement  robust joint AP clustering and beamforming design with imperfect CSI in cell-free  systems. Numerical results demonstrate that the proposed unsupervised deep learning algorithm achieves a higher worst-case sum rate under a smaller number of AP clustering with computational efficiency.
\end{abstract}


\section{Introduction}
\IEEEPARstart{R}{ecently}, cell-free systems have received significant attention\cite{bib17},\cite{bib42}.  By connecting all access points (APs) to a central processing unit (CPU) via backhaul links, cell-free systems allow multiple APs to simultaneously collaborate  to serve users within the network coverage area, which could  overcome many of the interference issues that appear in cellular networks \cite{bib41},\cite{bib12}. Nevertheless, popular beamforming design in cell-free systems generally assumes that all APs in the network coverage area serve users  simultaneously \cite{bib38},\cite{bib40}. This  appears to be impractical as long-range APs serving users consume precious power and bandwidth resources while contributing little useful power due to high path losses\cite{bib4}. 
To solve the above problem, a practical solution is to allow  a subset of APs in cell-free  systems to serve users simultaneously, which can also be called AP clustering. Consequently, joint AP clustering and beamforming design is proposed to improve both the sum rate performance and the practicality of cell-free systems. 

Theoretically, joint AP clustering and beamforming design belongs to the mixed-integer non-convex optimization problem that is difficult to solve efficiently. Some traditional optimization algorithms have approximated the solution of such mixed-integer non-convex optimization problem. In particular, based on block coordinate descent (BCD) \cite{bib35}, \cite{bib20} proposed a sparse weighted minimum mean square error (S-WMMSE) algorithm to consider joint AP clustering and beamforming design under the power constraint. \cite{bib34}  conducted joint user scheduling and beamforming design for multiuser multiple-input-multiple-output (MIMO) networks via fractional programming \cite{bib32} and Hungarian algorithm\cite{bib33}, where both user scheduling and AP clustering can be viewed as the integer programming problem.  By applying BCD \cite{bib35}, fractional
programming \cite{bib32} and compressive sensing \cite{bib36}, the work in \cite{bib37} optimized user scheduling, power allocation and beamforming  in cell-free systems. Although these traditional optimization algorithms could solve such mixed-integer non-convex optimization problem, they usually require multiple iterations and matrix inversions, thus imposing a severe computational burden.

In recent years, deep learning  has been widely deployed in wireless communications for improving communication performance and computational efficiency \cite{bib28}, \cite{bib22}.  In particular, a joint AP clustering and beamforming design based on unsupervised deep learning has been proposed in \cite{bib1}, where the convolutional neural networks (CNNs) mapped beamforming from channel state information (CSI) and an adaptive threshold ReLU (ATReLU) activation
function was added to the CNNs after to realize AP clustering. Since the CNNs and ATReLU activation function enable unsupervised end-to-end training, the work in \cite{bib1} optimizes AP  clustering and beamforming design simultaneously. Unfortunately, the ATReLU activation function is designed with only one AP clustering threshold between all APs and all users, which is difficult to achieve the optimal AP clustering results. The reason is that AP clustering is done in units of one AP to one user, which means that one AP clustering threshold is required between each AP and each user. As a result, it is necessary to carry out the research of unsupervised deep learning for joint AP clustering and beamforming design in cell-free  systems, where between each AP and each user is designed with an AP clustering threshold.

It is worth noting that beamforming design or AP clustering in most of these references, e.g., \cite{bib41}, \cite{bib38},\cite{bib40},\cite{bib20},\cite{bib22},\cite{bib1}, optimistically assumes the availability of perfect CSI, which leads to system performance degradation in  practice. To this end, it is highly desired to take the CSI  estimation errors  into account, and there have been some studies on robust beamforming design, especially for multicellular networks, e.g., in \cite{bib10},\cite{bib24},\cite{bib25}. However, these methods are generally solved by traditional optimization algorithms, where the computational complexity is  extremely high due to the need of multiple iterations and matrix inversions. Besides,  these methods  only consider the beamforming design without considering  AP clustering.

Based on the above considerations, in this paper, regarding to the
optimization problem of robust joint AP clustering and
beamforming design with  imperfect CSI in cell-free   systems, we propose a low computational complexity unsupervised deep learning algorithm  named as Robust Joint AP Clustering and Beamforming Network (RJAPCBN). The major contributions of this work are summarized as follows:
\begin{itemize}
	\item[1)] 
	An optimization model for robust joint AP clustering
	and beamforming design with imperfect CSI in cell-free systems is built, which aims at maximizing the worst-case sum rate and minimizing the number of AP clustering with imperfect CSI under the power constraint and the sparsity constraint for AP clustering simultaneously. By transformations, the intractable semi-infinite constraints in the optimization model caused by the imperfect CSI are converted into a more tractable form, paving the way for the design of a computationally efficient unsupervised deep learning algorithm. 
\end{itemize}
\begin{itemize}
	\item[2)] 
	The RJAPCBN is proposed to realize the mapping from CSI to beamforming with high computational efficiency. By designing an adaptive AP clustering module, the proposed RJAPCBN  also ensures that the output beamforming satisfies the sparsity constraint of AP clustering. In addition, the adaptive AP clustering module also proposes a differentiable threshold function to ensure that an AP clustering threshold between each AP and each user is adaptively set, which effectively reduces the impractical drawback of cell-free  systems, i.e., longer-range APs serving users consume valuable power and bandwidth	resources while contributing little useful power due to	high path losses. 
\end{itemize}

\begin{itemize}
	\item[3)] 
	Numerical results are conducted to validate the effectiveness of the proposed RJAPCBN. In terms of performance, the proposed RJAPCBN achieves a higher worst-case sum rate under a smaller number of AP clustering. As for  computational complexity, the number	of real multiplication of the proposed RJAPCBN is about $10^6$, which is much lower than other traditional and deep learning algorithms such as S-WMMSE\cite{bib20}, WMMSE\cite{bib23} and CNNs \cite{bib22}. 
\end{itemize}

The rest of this paper is organized as follows: In Section \uppercase\expandafter{\romannumeral2}, the system model and optimization problem are introduced. In Section \uppercase\expandafter{\romannumeral3}, the optimization problem is transformed into a more tractable form. In Section \uppercase\expandafter{\romannumeral4}, the computationally effective unsupervised deep learning RJAPCBN is proposed. Finally, numerical results and
conclusions are provided in Sections \uppercase\expandafter{\romannumeral5}, and \uppercase\expandafter{\romannumeral6}, respectively.

\textit{Notations:} The scalars, vectors, and matrices are denoted
by lowercase letter $x$, boldface lowercase letter $\mathbf{x}$, and boldface uppercase letter $\mathbf{X}$, respectively. $\mathbb{C}$ and $\mathbb{R}$ denotes the set of complex and real numbers, respectively. $(\cdot)^H$ denotes the conjugate transpose. $\left|\cdot\right| $, $\left\Vert \cdot\right\Vert_1 $  and $\left\Vert \cdot\right\Vert_2 $ denote the modulus of complex numbers, $\ell_{1}$ and  $\ell_{2}$ norms, respectively. $\operatorname{Re}\left\{\cdot\right\}$ denotes the real part of complex numbers. $\mathbf{A}\succeq0$ denotes that $\mathbf{A}$ is a semi-positive definite matrix.

\section{System Model and Problem Formulation}
\subsection{System Model}
Consider a downlink cell-free system with $Q$ APs and $I$ single-antenna users, where AP is equipped with $M$ antennas. All APs are connected to a CPU via backhaul links, in which the CPU makes resource allocation decisions for all APs.  Let $\mathcal{Q}=\left\{1, \cdots, Q\right\}$ and $\mathcal{I} =\left\{1, \cdots, I\right\}$ denote the sets of APs and users, respectively. To simplify the notation, $ i $ and $ j $ denote the user's indexes, and  $ q $  denotes the AP's index. The received signal of the $i^{th}$ user is denoted as
\begin{equation}
	{y}_{i}=\mathbf{{h}}_{i}^H \mathbf{v}_{i} {s}_{i}+\sum_{j \neq i} \mathbf{{h}}_{i}^H \mathbf{v}_{j} {s}_{j}+{z}_{i}, \forall i,j\in\mathcal{I},
	\label{eq1}
\end{equation}
where ${s}_{i}$ denotes the data being sent to the $ i^{th} $ user. ${z}_{i}$ denotes the additive noise following
the complex Gaussian distribution $\mathcal{C} \mathcal{N} \left(0, \sigma_{i}^{2} \right)$. $\mathbf{h}_{i}=\left[\begin{array}{lllll}\mathbf{h}_{i}^{1,H},  \cdots , \mathbf{{h}}_{i}^{q,H}, \cdots , \mathbf{{h}}_{i}^{Q,H}\end{array}\right]^H \in \mathbb{C}^{QM \times 1}$ denotes the CSI of the AP set $\mathcal{Q}$ to the $ i^{th} $ user, in which $ \mathbf{{h}}_{i}^{q}  \in \mathbb{C}^{M \times 1} $ is the CSI of the $ q^{th} $ AP to the $ i^{th} $ user. $\mathbf{v}_{i}=\left[\mathbf{v}_{i}^{1,H}, \cdots, \mathbf{v}_{i}^{q,H}, \cdots, \mathbf{v}_{i}^{Q,H}\right]^{{H}} \in \mathbb{C}^{QM \times 1}$ denotes the beamforming of the AP set $\mathcal{Q}$ to the $ i^{th} $ user, and $\mathbf{v}_{i}^{q} \in \mathbb{C}^{M \times 1}$ is the beamforming of the $ q^{th}$ AP to the $ i^{th} $ user. 

From Eq.(\ref{eq1}), the signal-to-interference-plus-noise ratio (SINR) of the $i^{th}$ user is expressed as 
\begin{equation}
	\text{SINR}_{i}=\frac{\left| \mathbf{{h}}_{i}^H \mathbf{v}_{i}\right|^2 }{\sum\limits_{j \neq i} \left| \mathbf{{h}}_{i}^H \mathbf{v}_{j}\right|^2 +\sigma_{i}^{2}},\forall i,j\in\mathcal{I},
	\label{eq2}
\end{equation}
so that the achievable rate  of  the $i^{th}$ user is
\begin{equation}
	{R}_{i}=\log_2 \left(1+	\text{SINR}_{i}\right),\forall i\in\mathcal{I}.
	\label{eq3}
\end{equation}

\subsection{AP Clustering}
Cell-free systems typically  assume all APs in the network coverage area serving users simultaneously, which seems impractical\cite{bib4}.  In this regard, it is encouraged to select a subset of APs  $\mathcal{S}_i\subseteq \mathcal{Q}$ to serve the $ i^{th} $  user, i.e., when the  $ i^{th} $ user is not served by the $ q^{th} $ AP, the beamforming $\mathbf{v}_{i}^{q} $  can be set to zero, and vice verse. Formally, this is denoted as
\begin{equation}
	\mathbf{v}_{i}^{q}=\mathbf{\textbf{0}},\forall q\notin\mathcal{S}_i,\forall i\in\mathcal{I}.
	\label{eq4}
\end{equation} 

For AP clustering, it is expected that a smaller subset of APs $\mathcal{S}_i$ serves the  $ i^{th} $ user, i.e., the beamforming $\mathbf{v}_{i}$  should be a sparse structure containing a larger number of zero blocks\cite{bib1}. A popular way to enforce the sparsity of the solution to an optimization problem uses a norm to penalize the objective function such as the $ \ell_{1}$ norm\cite{bib3}. Consequently, by jointly considering AP clustering and  beamforming design, the optimization objective can be defined as 
\begin{equation}
S_{\text{spa}}=\underset{i\in\mathcal{I}}{\sum}\big(R_{i}-\lambda\underset{q\in \mathcal{Q}}{\sum}\left\Vert \mathbf{v}_{i}^{q}\right\Vert_1\big),
\label{eq5}
\end{equation}
where $S_{\text{spa}}$ denotes a penalized sparse
sum rate. Note that maximizing $S_{\text{spa}}$ enables the goal of maximizing the sum rate and minimizing the number of AP clustering simultaneously. $\lambda\geq0$ denotes the parameter balancing the sum rate and the number of AP clustering.

\subsection{CSI Error Model}

It is well known that the perfect CSI assumption generally leads to the inevitable loss of system performance in  practice \cite{bib12}. To this end, this work applies a bounded model to characterize CSI estimation errors \cite{bib10}, since it is able to capture different types of CSI errors, e.g., the errors caused by noise, quantization, finite feedback, etc. Specifically, let $\Delta\mathbf{h}_{i}^{q}$ denote the estimation errors of $\mathbf{h}_{i}^{q}$, and then the actual CSI can be denoted as a combination of the estimated CSI and the corresponding estimation errors, i.e.,
\begin{equation}
	\mathbf{h}_{i}^{q}=\hat{\mathbf{h}}_{i}^{q}+\Delta\mathbf{h}_{i}^{q},\left\|\Delta\mathbf{h}_{i}^{q}\right\|_{2} \leq \epsilon_i^q,\forall i\in\mathcal{I},\forall q\in\mathcal{Q},
	\label{eq6}
\end{equation}
where $\hat{\mathbf{h}}_{i}^{q}$ denotes the estimated CSI of the $ q^{th} $ AP to the $ i^{th} $ user, and $\left\|\Delta\mathbf{h}_{i}^{q}\right\|_{2} \leq \epsilon_i^q$ denotes that  $\Delta\mathbf{h}_{i}^{q}$ is limited within an origin hyperspherical region of radius  $\epsilon_i^q$ \cite{bib12}, \cite{bib10}.

\subsection{Problem Formulation}
In this paper, our goal is to achieve a robust joint AP clustering and beamforming design in cell-free systems by maximizing the worst-case penalized sparse sum rate of all users with imperfect CSI, taking into account the power constraint and the sparsity constraint for AP clustering. The optimization problem is formulated as
\begin{equation}
	\begin{split}
		\underset{\mathbf{v}_{i}^{q}}{\textrm{max}}\,\,\,\underset{\Delta\mathbf{h}_{i}^{q}}{\textrm{min}}&\,\,\,S_{\text{spa}}
		\\\textrm{s.t.}&\,\,\,
		\textrm{C1}:\,\,\underset{i\in\mathcal{I}}{\sum}\left(\mathbf{v}_{i}^{q}\right)^{{H}}\mathbf{v}_{i}^{q}\leq \textrm{P}_{\textrm{max}} ,\forall q\in\mathcal{Q},
		\\&\,\,\, \textrm{C2}:\,\,\mathbf{v}_{i}^{q}=\mathbf{\textbf{0}},\forall q\notin\mathcal{S}_i,\forall i\in\mathcal{I},
	\end{split}
	\label{eq7}
\end{equation}
where $\textrm{P}_{\textrm{max}}$ denotes the AP maximum power. For the optimization problem (\ref{eq7}), the optimization objective is non-convex and non-smooth, in which \textrm{C2} is an integer constraint. This means that the optimization problem (\ref{eq7}) is a mixed integer nonconvex and nonsmooth optimization problem, which is challenging to solve.

%\section{Proposed Method}
%Due to the intractable form of the optimization problem (\ref{eq7}), we first propose a transformation to recast the optimization problem (\ref{eq7}) to a more tractable form, which paves the way for designing an effective unsupervised deep learning algorithm. Afterwards, we propose a low computational complexity unsupervised deep learning algorithm RJAPCBN to achieve  the goal of  robust joint AP clustering and beamforming design with imperfect CSI in cell-free  systems.

\section{Problem Transformation}
In this section, we provide a useful transformation to simplify the optimization problem  in (\ref{eq7}). Concretely, to make the optimization problem (\ref{eq7}) more tractable, we introduce a slack variable $\boldsymbol{{\gamma}}=\left\{\gamma_1, \cdots, \gamma_i,\cdots,\gamma_	I\right\}$ to replace the worst-case SINR terms, so that the optimization problem (\ref{eq7}) can be rewritten as
\begin{equation}
	\begin{split}
		\underset{\mathbf{v}_{i}^{q},\gamma_i}{\textrm{max}}&\,\,\, \underset{i\in\mathcal{I}}{\sum}\big(\log_2 \left(1+\gamma_{i}\right)-\lambda\underset{q\in \mathcal{Q}}{\sum}\left\Vert \mathbf{v}_{i}^{q}\right\Vert_1\big)
		\\\textrm{s.t.}&\,\,\,
		\textrm{C1},\textrm{C2},
		\\&\,\,\, \textrm{C3}:\,\,\underset{\Delta\mathbf{h}_{i}^{q}}{\textrm{min}}\,\, \text{SINR}_{i}\geq\gamma_{i},\forall i\in\mathcal{I},
	\end{split}
	\label{eq8}
\end{equation}
where \textrm{C3} contains  the intractable semi-infinite constraints caused by the imperfect CSI $\left\|\Delta\mathbf{h}_{i}^{q}\right\|_{2} \leq \epsilon_i^q,\forall i\in\mathcal{I},\forall q\in\mathcal{Q}$. In the following, we transform it into a more tractable form.

To begin with, it is obvious that the lower bound for the worst-case SINR  of the $i^{th}$ user is obtained as
\begin{equation}
	\underset{\Delta\mathbf{h}_{i}^{q}}{\textrm{min}} \,\,\text{SINR}_{i}\geq\frac{\underset{\Delta\mathbf{h}_{i}^{q}}{\textrm{min}} \,\,\left| \mathbf{{h}}_{i}^H \mathbf{v}_{i}\right|^2 }{\underset{\Delta\mathbf{h}_{i}^{q}}{\textrm{max}} \,\,\sum\limits_{j \neq i} \left| \mathbf{{h}}_{i}^H \mathbf{v}_{j}\right|^2 +\sigma_{i}^{2}},\forall i,j\in\mathcal{I}.
	\label{eq9}
\end{equation}
To achieve the goal of robust design, \textrm{C3} can be replaced with this lower bound, thus a lower bound performance of the original optimization problem can be obtained. Accordingly, this leads to $	\overline{\textrm{C3}}$
\begin{equation}
	\overline{\textrm{C3}}:\,\,\frac{\underset{\Delta\mathbf{h}_{i}^{q}}{\textrm{min}} \,\,\left| \mathbf{{h}}_{i}^H \mathbf{v}_{i}\right|^2 }{\underset{\Delta\mathbf{h}_{i}^{q}}{\textrm{max}} \,\,\sum\limits_{j \neq i} \left| \mathbf{{h}}_{i}^H \mathbf{v}_{j}\right|^2 +\sigma_{i}^{2}} \geq\gamma_{i},\forall i,j\in\mathcal{I}.
	\label{eq10}
\end{equation}

Subsequently, to facilitate the solution of  $\overline{\textrm{C3}}$, it is transformed into the following constraints by introducing the slack variables, i.e.,
\begin{equation}
{\textrm{C4}}:\,\,\underset{\Delta\mathbf{h}_{i}^{q}}{\textrm{min}}\,\,\left| \mathbf{{h}}_{i}^H \mathbf{v}_{i}\right|^2\geq\alpha_{i},\forall i\in\mathcal{I},
\label{eq11}
\end{equation}
\begin{equation}
	\textrm{C5}:\,\,\underset{\Delta\mathbf{h}_{i}^{q}}{\textrm{max}}\,\,\sum\limits_{j \neq i} \left| \mathbf{{h}}_{i}^H \mathbf{v}_{j}\right|^2 +\sigma_{i}^{2} \leq \beta_{i},\forall i,j\in\mathcal{I},
	\label{eq12}
\end{equation}
\begin{equation}
{\textrm{C6}}:\,\,\frac{\alpha_{i} }{\beta_{i}} \geq\gamma_{i},\forall i\in\mathcal{I},
	\label{eq13}
\end{equation}
where $\boldsymbol{{\alpha}}=\left\{\alpha_1, \cdots, \alpha_i,\cdots,\alpha_	I\right\}$ and $\boldsymbol{{\beta}}=\left\{\beta_1, \cdots, \beta_i,\cdots,\beta_I\right\}$ are the slack
variables to decompose the fractions. Although \textrm{C6}  is a simple convex constraint, \textrm{C5}  and \textrm{C6}  are still the intractable semi-infinite constraints due to imperfect CSI. On the other hand, the S-procedure \cite{bib11} is an efficient transformation technique to convert the intractable semi-infinite constraints into the tractable forms of the linear matrix inequality. As a result, this work applies the S-procedure to transform  \textrm{C4} into the linear matrix inequality with the following \emph{Lemma 1}.

\emph{Lemma 1: (S-Procedure \cite{bib11}) Define the quadratic functions of the variable $ \mathbf{x}\in \mathbb{C}^{N\times 1}$:
$$
f_{k}(\mathbf{x})=\mathbf{x}^{{H}} \mathbf{A}_{k} \mathbf{x}+2\operatorname{Re}\left\{\mathbf{a}_{k}^{{H}} \mathbf{x}\right\}+a_{k}, k=0, 1,
$$
where $\mathbf{A}_{k}=\mathbf{A}_{k}^H\in \mathbb{C}^{N\times N} $, $\mathbf{a}_{k}\in \mathbb{C}^{N\times 1}$ and $a_{k}\in \mathbb{C}^{1\times 1}$. The condition $f_{1}(\mathbf{x}) \geq 0 \Rightarrow f_{0}(\mathbf{x}) \geq 0$ hold if and only if there exist $\delta\geq0 $ such that 
$$
\left[\begin{array}{cc}\mathbf{A}_{0} & \mathbf{a}_{0} \\ \mathbf{a}_{0}^{{H}} & a_{0}\end{array}\right]- \delta\left[\begin{array}{cc}\mathbf{A}_{1} & \mathbf{a}_{1} \\ \mathbf{a}_{1}^{{H}} & a_{1}\end{array}\right] \succeq \mathbf{0}.
$$}

By bringing $\mathbf{h}_{i}^{q}=\hat{\mathbf{h}}_{i}^{q}+\Delta\mathbf{h}_{i}^{q}$ in Eq.(\ref{eq6}) into Eq.(\ref{eq11}),  \textrm{C4} is  rewritten as
\begin{equation}
	\begin{aligned}
	\underset{\Delta\mathbf{h}_{i}^{q}}{\textrm{min}}\,\,&
	\Delta\mathbf{h}_{i}^{H}\mathbf{E}_{i,i}\Delta\mathbf{h}_{i}+2\operatorname{Re}\left\{\mathbf{e}_{i,i}^{{H}} \Delta\mathbf{h}_{i}\right\}\\&+\hat{\mathbf{h}}_{i}^{H}\mathbf{E}_{i,i}\hat{\mathbf{h}}_{i}-\alpha_{i}\geq0,\forall i\in\mathcal{I}
	\label{eq14}
\end{aligned}
\end{equation}
with $\hat{\mathbf{h}_{i}}=\left[\begin{array}{lllll}\hat{\mathbf{h}}_{i}^{1,H},  \cdots , \hat{\mathbf{{h}}}_{i}^{q,H}, \cdots , \hat{\mathbf{{h}}}_{i}^{Q,H}\end{array}\right]^H \in \mathbb{C}^{QM \times 1}$, $\Delta\mathbf{h}_{i}=\left[\begin{array}{lllll}\Delta\mathbf{h}_{i}^{1,H},  \cdots , \Delta\mathbf{{h}}_{i}^{q,H}, \cdots , \Delta\mathbf{{h}}_{i}^{Q,H}\end{array}\right]^H \in \mathbb{C}^{QM \times 1}$, $\mathbf{E}_{i,i}=\mathbf{v}_{i}\mathbf{v}_{i}^H\in \mathbb{C}^{QM \times QM}$ and  $\mathbf{e}_{i,i}=\mathbf{v}_{i}\mathbf{v}_{i}^H \hat{\mathbf{h}_{i}}\in \mathbb{C}^{QM \times 1}$. According to  \emph{Lemma 1}, \textrm{C4} is transformed into the linear matrix inequality with the new introduced slack variables $\boldsymbol{{\delta}}=\left\{\delta_1, \cdots, \delta_i,\cdots,\delta_I\right\}$, which is denoted as 
\begin{equation}
\begin{aligned}
\overline{\textrm{C4}}:\left[\begin{array}{cc}\mathbf{E}_{i,i}+\delta_i\mathbf{I} & \mathbf{e}_{i,i} \\ \mathbf{e}_{i,i}^{{H}} & \hat{\mathbf{h}}_{i}^{H}\mathbf{E}_{i,i}\hat{\mathbf{h}}_{i}-\alpha_{i}-\delta_i\epsilon_i^2\end{array}\right] \succeq \mathbf{0},\forall i\in\mathcal{I},
\label{eq15}
\end{aligned}
\end{equation}
where  $\epsilon_{i}\triangleq \sqrt{\sum_{q=1}^{Q} (\epsilon_i^q)^{2}}$ based on Eq.(\ref{eq6}) and \cite{bib12}.

Afterwards, the semi-infinite constraints in \textrm{C5} is transformed into the linear matrix inequality. Due to the summation term in \textrm{C5}, this is difficult to directly apply the S-procedure in \emph{Lemma 1} to transform into the linear matrix inequality. For this reason, this work applies the sign-definiteness \cite{bib13} that can be regarded as an extended version of the S-procedure to transform  \textrm{C5} into the linear matrix inequality with the following \emph{Lemma 2}.

\emph{Lemma 2: (Sign-Definiteness \cite{bib13}) For a given set of matrices $\mathbf{A}=\mathbf{A}^H $, $\mathbf{Y}$ and $\mathbf{Z}$, the follow linear matrix inequality meets
	$$
	\mathbf{A} \succeq \mathbf{Y}^{{H}} \mathbf{X} \mathbf{Z}+\mathbf{Z}^{{H}} \mathbf{X}^{{H}} \mathbf{Y},\left\|\mathbf{X}\right\|_{2} \leq \epsilon,
	$$
	if and only if there exist real numbers $\mu\geq0 $  such that 	
	$$
	\left[\begin{array}{cccc}\mathbf{A} -\mu \mathbf{Z}^{{H}} \mathbf{Z} & -\epsilon \mathbf{Y}^{{H}} \\ -\epsilon \mathbf{Y} & \mu \mathbf{I} \end{array}\right] \succeq \mathbf{0}.
	$$}
	
Defining $\mathbf{V}=\left[\mathbf{v}_1, \cdots,\mathbf{v}_{i-1}, \mathbf{v}_{i+1},\cdots,\mathbf{v}_I\right]\in \mathbb{C}^{QM \times (I-1)}$, \textrm{C5} is first equivalently converted into a matrix inequality by utilizing the Schur's complement lemma \cite{bib14}, i.e., 
\begin{equation}
	\underset{\Delta\mathbf{h}_{i}^{q}}{\textrm{min}}\,\,\left[\begin{array}{cc}\beta_{i}-\sigma_{i}^{2} & \mathbf{{h}}_{i}^H \mathbf{V}  \\ \mathbf{V}^H \mathbf{{h}}_{i}& \mathbf{I}_{(I-1)}\end{array}\right] \succeq \mathbf{0}, \forall i\in\mathcal{I}
	\label{eq16}
\end{equation}
with $\mathbf{h}_{i}=\hat{\mathbf{h}}_{i}+\Delta\mathbf{h}_{i}$, then Eq.(\ref{eq16}) is rewritten as
\begin{equation}
	\begin{aligned}
	\underset{\Delta\mathbf{h}_{i}^{q}}{\textrm{min}}\,&\left[\begin{array}{cc}\beta_{i}-\sigma_{i}^{2} & \hat{\mathbf{h}}_{i}^{H} \mathbf{V}\\ \mathbf{V}^H \hat{\mathbf{h}}_{i}  & \mathbf{I}_{(I-1)}\end{array}\right] \succeq\\&-\bigg(\left[\begin{array}{cc} \mathbf{0}_{1\times QM}\\\mathbf{V}^H   \end{array}\right] \Delta\mathbf{h}_{i}\left[\begin{array}{cc}  1 &\mathbf{0}_{1\times(I-1)} \end{array}\right]
	\\&+\left[\begin{array}{cc} 1\\   \mathbf{0}_{(I-1)\times1}\end{array}\right] \Delta\mathbf{h}_{i}^H\left[\begin{array}{cc}  \mathbf{0}_{QM\times 1}&\mathbf{V} \end{array}\right]\bigg),
	\forall i\in\mathcal{I}.
	\label{eq17}
\end{aligned}
\end{equation}

According to  \emph{Lemma 2}, \textrm{C5} is transformed into the linear matrix inequality with the new introduced slack variables $\boldsymbol{{\mu}}=\left\{\mu_1, \cdots, \mu_i,\cdots,\mu_I\right\}$, which is denoted as
\begin{equation}
	\overline{\textrm{C5}}:\left[\begin{array}{ccc}\beta_{i}-\sigma_{i}^{2}-\mu_i & \hat{\mathbf{h}}_{i}^{H} \mathbf{V} & \mathbf{0}_{1\times QM} \\ \mathbf{V}^H \hat{\mathbf{h}}_{i} & \mathbf{I}_{(I-1)} & \epsilon_{i} \mathbf{V}^H \\ \mathbf{0}_{QM \times 1} & \epsilon_{i} \mathbf{V}  & \mu_{i} \mathbf{I}_{QM}\end{array}\right] \succeq \mathbf{0}, \forall i \in \mathcal{I}.
	\label{eq18}
\end{equation}

Based on the above transformation, the original optimization problem (\ref{eq7}) is recast as
\begin{equation}
	\begin{split}
		\underset{\mathbf{v}_{i}^{q},\alpha_i,\beta_i,\gamma_i,\delta_i,\mu_i}{\textrm{max}}&\,\, \underset{i\in\mathcal{I}}{\sum}\big(\log_2 \left(1+\gamma_{i}\right)-\lambda\underset{q\in \mathcal{Q}}{\sum}\left\Vert \mathbf{v}_{i}^{q}\right\Vert_1\big)
		\\\textrm{s.t.}&\quad
		\textrm{C1},\textrm{C2},\overline{\textrm{C4}},\overline{\textrm{C5}},\textrm{C6}.
	\end{split}
	\label{eq19}
\end{equation}
Note that the optimization problem (\ref{eq19}) has transformed the semi-infinite constraints caused by the imperfect CSI into the linear matrix inequality  with the slack variables.

\section{Proposed  RJAPCBN}
\begin{figure*}[t]
	\centering
	\includegraphics[width=15.7cm,height=15.7cm]{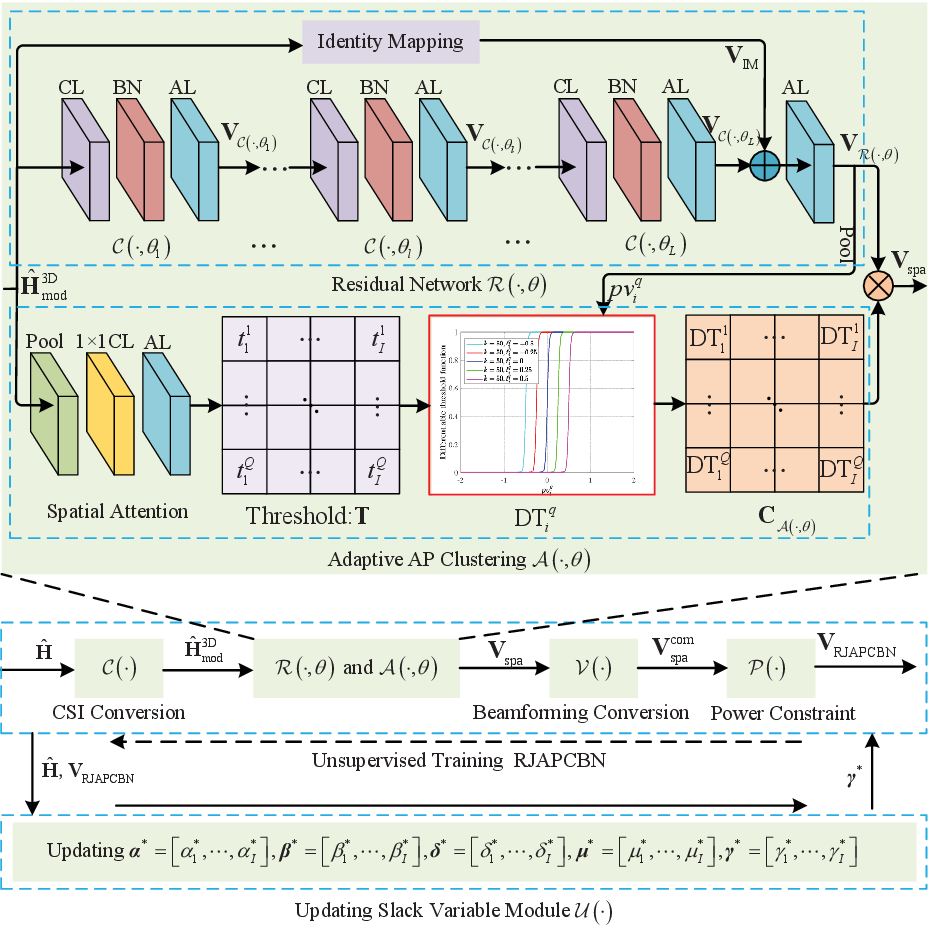}
	\caption{The model architecture of the proposed  RJAPCBN.}
\end{figure*}

In this section, we pay our attention on designing a computationally effective unsupervised deep learning method RJAPCBN to achieve robust joint AP clustering and beamforming design with imperfect CSI in cell-free systems by solving the optimization problem (\ref{eq19}). As illustrated in Fig.1, the  proposed RJAPCBN first designs the CSI conversion $\mathcal{C}\left( \cdot\right) $, residual network $\mathcal{R}\left( \cdot,\theta\right) $, adaptive AP clustering $\mathcal{A}\left( \cdot,\theta\right) $, beamforming conversion $\mathcal{V}\left( \cdot\right) $ and power constraint $\mathcal{P}\left( \cdot\right) $ to output a sparse beamforming $\mathbf{V}_{\text{RJAPCBN}}$ that satisfies both  $\textrm{C1}$ and $\textrm{C2}$ in the optimization problem (\ref{eq19}) by taking $\hat{\mathbf{H}}=\left[\hat{\mathbf{h}}_{1}, \cdots, \hat{\mathbf{h}}_{i}, \cdots, \hat{\mathbf{h}}_{I}\right] \in \mathbb{C}^{QM \times I}$ as an input. Subsequently, with  $\mathbf{V}_{\text{RJAPCBN}}$ and $\hat{\mathbf{H}}$, the updating slack variable module $\mathcal{U}\left( \cdot\right) $  in the proposed RJAPCBN updates the slack variables $\boldsymbol{{\alpha}}$, $\boldsymbol{{\beta}}$, $\boldsymbol{{\delta}}$, $\boldsymbol{{\mu}}$ and $\boldsymbol{{\gamma}}$ by simultaneously satisfying $\overline{\textrm{C4}}$, $\overline{\textrm{C5}}$ and $\textrm{C6}$ in the optimization problem (\ref{eq19}). Finally, based on the obtained $\boldsymbol{{\gamma}}$, the proposed  RJAPCBN is unsupervised trained  with the negative of the objective function of the optimization problem (\ref{eq19}) as the loss function.

\subsection{CSI Conversion $\mathcal{C}\left( \cdot\right) $}
The CSI of communication systems is complex numbers, while deep learning algorithms such as CNNs usually deal with three-dimensional (3D) real numbers. For this purpose, $\mathcal{C}\left( \cdot\right) $ transforms the estimated two-dimensional (2D) complex CSI  $\hat{\mathbf{H}}=\left[\hat{\mathbf{h}}_{1}, \cdots, \hat{\mathbf{h}}_{i}, \cdots, \hat{\mathbf{h}}_{I}\right] \in \mathbb{C}^{QM \times I}$ of the AP set $\mathcal{Q}$ to the user set $\mathcal{I}$ into a 3D real CSI. Specifically, $\hat{\mathbf{H}}$ is computed with the modulus value to obtain a 2D real CSI $\hat{\mathbf{H}}_{\text{mod}}^{\text{2D}}\in \mathbb{R}^{QM \times I}$. Subsequently, $\hat{\mathbf{H}}_{\text{mod}}^{\text{2D}}\in \mathbb{R}^{QM \times I}$ is transformed into a 3D real CSI $\hat{\mathbf{H}}_{\text{mod}}^{\text{3D}}\in \mathbb{R}^{Q \times I \times M}$.\footnote{In this paper, the first, second and third dimensions of a 3D tensor are denoted as width, height and third dimension, respectively.}

\subsection{Residual Network $\mathcal{R}\left(\cdot,\theta \right) $} $\mathcal{R}\left(\cdot,\theta \right) $ achieves the mapping from 3D real CSI $\hat{\mathbf{H}}_{\text{mod}}^{\text{3D}}\in \mathbb{R}^{Q \times I \times M}$ to beamforming. As the unique weight sharing mechanism of the CNNs significantly reduces the computational complexity of neural networks, this is in line with the goal of designing a low computational complexity unsupervised deep learning algorithm. Therefore, $\mathcal{R}\left(\cdot,\theta \right) $ selects the CNNs to achieve the mapping from $\mathbf{H}_{\text{mod}}^{\text{3D}}\in \mathbb{R}^{Q \times I \times M}$ to beamforming. To be specific, $\mathcal{R}\left(\cdot,\theta \right) $ contains $L$ layers, where each layer contains a convolution unit with a convolution layer (CL), batch normalization (BN) layer, activation layer (AL). Formally,  for the $l^{th}$ layer, denoted as  $\mathcal{C}\left(\cdot,\theta_l \right) $, its formula is defined as
\begin{equation}
	\mathbf{V}_{\mathcal{C}\left(\cdot,\theta_l \right)}=\text{AL}\left(\text{BN}\left(\text{CL}\left(\mathbf{V}_{{\mathcal{C}\left(\cdot,\theta_{l-1} \right)}}, \theta_l\right)\right)\right),
	\label{eq20}
\end{equation}
where $\mathbf{V}_{\mathcal{C}\left(\cdot,\theta_l \right)}$ denotes the output of $\mathcal{C}\left(\cdot,\theta_l \right) $, and $\theta_l$ is the parameters of $\mathcal{C}\left(\cdot,\theta_l \right) $. $\mathbf{V}_{{\mathcal{C}\left(\cdot,\theta_{l-1} \right)}}$ denotes the input of $\mathcal{C}\left(\cdot,\theta_l \right) $, note that $\mathbf{V}_{{\mathcal{C}\left(\cdot,\theta_{0} \right)}}=\hat{\mathbf{H}}_{\text{mod}}^{\text{3D}}\in \mathbb{R}^{Q \times I \times M}$. $\text{CL}\left(\cdot,\cdot \right) $ denotes the convolution operation. $\text{BN}\left(\cdot \right) $ denotes the BN operation, which is
usually added after the CL to reduce the overfitting probability\cite{bib5}.  $\text{AL}\left(\cdot \right) $ denotes the AL operation, which selects the commonly used ReLU activation function  $ \text{ReLU}(x)=\max (0, x) $ to implement nonlinear operations\cite{bib6}. Note that the last layer of $\mathcal{R}\left(\cdot,\theta \right) $, i.e., $\mathcal{C}\left(\cdot,\theta_L \right) $, outputs the real and imaginary parts of  beamforming, which should contain both positive and negative values. Consequently, $\text{AL}\left(\cdot \right) $ in $\mathcal{C}\left(\cdot,\theta_L \right) $  can adopt the Tanh activation function $\text{Tanh}(x)=\frac{e^{x}-e^{-x}}{e^{x}+e^{-x}}$.

In addition,  the residual structure of the CNNs can effectively avoid the gradient disappearance problem. As a result, following \cite{bib7}, $\mathcal{R}\left(\cdot,\theta \right) $ adds an identity mapping on top of $\mathcal{C}\left(\cdot,\theta_l \right), l=1,\cdots, L $ to construct the residual structure. Such that the output of $\mathcal{R}\left(\cdot,\theta \right) $ is denoted as:
\begin{equation}
	\mathbf{V}_{\mathcal{R}\left(\cdot,\theta \right)}=\text{AL}\left(\mathbf{V}_{\mathcal{C}\left(\cdot,\theta_L \right)}+\mathbf{V}_{\text{IM}} \right),
	\label{eq21}
\end{equation} 
where $\mathbf{V}_{\text{IM}}$  denotes the output of the  identity  mapping utilizing  a $1\times 1$ CL with   $\hat{\mathbf{H}}_{\text{mod}}^{\text{3D}}\in \mathbb{R}^{Q \times I \times M}$ as the output. Note that the architectural parameters of the $1\times 1$ CL in the identity  mapping are adjusted to ensure that $\mathbf{V}_{\text{IM}}$ and $\mathbf{V}_{\mathcal{C}\left(\cdot,\theta_L \right)}$ have the same dimension, in which $\mathbf{V}_{\text{IM}}$ and $\mathbf{V}_{\mathcal{C}\left(\cdot,\theta_L \right)}$  are added to form the residual structure to avoid the gradient disappearance problem \cite{bib7}.  

\emph{Remark 1: For optimization problem (\ref{eq19}), the beamforming in cell-free systems has the following properties. When the dimension of the input 3D real CSI $\hat{\mathbf{H}}_{\text{mod}}^{\text{3D}}$ is $Q\times I \times M$, the dimension of the output beamforming should be a 3D complex tensor of dimension $Q\times I \times M$, which can be transformed into a 3D real tensor of dimension $Q\times I \times 2M$.}

Based on \emph{Remark 1}, when  $\hat{\mathbf{H}}_{\text{mod}}^{\text{3D}}\in \mathbb{R}^{Q \times I \times M}$ is inputted to $\mathcal{R}\left(\cdot,\theta \right) $, the dimension of the output beamforming $\mathbf{V}_{\mathcal{R}\left(\cdot,\theta \right)}$ should be $Q\times I \times 2M$. However, the dimension of $\mathbf{V}_{\mathcal{R}\left(\cdot,\theta \right)}$ is determined by the architectural parameters of the CL in $\mathcal{R}\left(\cdot,\theta \right) $ such as the convolution kernel size, convolution kernel number, sliding step size and zero padding size. Consequently, in what follows, we derive some architectural conditions for the CL in $\mathcal{R}\left(\cdot,\theta \right)$ to satisfy the dimension of $\mathbf{V}_{\mathcal{R}\left(\cdot,\theta \right)}$ as $Q\times I \times 2M$ when $\hat{\mathbf{H}}_{\text{mod}}^{\text{3D}}\in \mathbb{R}^{Q \times I \times M}$ is inputted.

\emph{Proposition 1: Let $w_{\mathcal{C}\left(\cdot,\theta_l \right)}^{\text{in}}$,  $h_{\mathcal{C}\left(\cdot,\theta_l \right)}^{\text{in}}$, $w_{\mathcal{C}\left(\cdot,\theta_l \right)}^{\text{out}}$ and $h_{\mathcal{C}\left(\cdot,\theta_l \right)}^{\text{out}}$ denote the input and output width and height dimensions of $\mathcal{C}\left(\cdot,\theta_l \right) $ in $\mathcal{R}\left(\cdot,\theta \right) $, as well as $k_l^w$, $k_l^h$, $p_l^w$, $p_l^h$, $s_l^w$ and $s_l^h$ denote the width and height dimensions of the convolution kernel, zero padding, sliding step for the CL of $\mathcal{C}\left(\cdot,\theta_l \right) $ in $\mathcal{R}\left(\cdot,\theta \right) $, respectively. When $s_l^w=1$, $s_l^h=1$, if $p_l^w=\frac{1}{2}(k_l^w-1)$, $p_l^h=\frac{1}{2}(k_l^h-1)$, both $p_l^w$,  $p_l^h$ and $k_l^w$, $k_l^h$ are positive integers, then $w_{\mathcal{C}\left(\cdot,\theta_l \right)}^{\text{out}}=w_{\mathcal{C}\left(\cdot,\theta_l \right)}^{\text{in}}$ and  $h_{\mathcal{C}\left(\cdot,\theta_l \right)}^{\text{out}}=h_{\mathcal{C}\left(\cdot,\theta_l \right)}^{\text{in}}$.}

\emph{Proof:} As can be seen in Fig.1, $\mathcal{C}\left(\cdot,\theta_l \right) $ includes one CL, BN and AL. For the CL in $\mathcal{C}\left(\cdot,\theta_l \right) $, its output width and height dimensions $w_{\mathcal{C}\left(\cdot,\theta_l \right)}^{\text{CL}} \times h_{\mathcal{C}\left(\cdot,\theta_l \right)}^{\text{CL}} $ are denoted as
\begin{equation}
	\begin{cases}
		w_{\mathcal{C}\left(\cdot,\theta_l \right)}^{\text{CL}}=\frac{w_{\mathcal{C}\left(\cdot,\theta_l \right)}^{\text{in}}+2 p_{l}^w-k_{l}^w}{s_l^w}+1,\\ 
		h_{\mathcal{C}\left(\cdot,\theta_l \right)}^{\text{CL}}\thinspace=\frac{h_{\mathcal{C}\left(\cdot,\theta_l \right)}^{\text{in}}\thinspace+2 p_l^h-k_{l}^h\thinspace}{s_l^h}+1,
		\label{eq31}
	\end{cases}
\end{equation}
where $s_l^w=1$, $s_l^h=1$, $p_l^w=\frac{1}{2}(k_l^w-1)$ and $p_l^h=\frac{1}{2}(k_l^h-1)$ are brought into Eq.(\ref{eq31}), i.e.,
\begin{equation}
	\begin{cases}
		w_{\mathcal{C}\left(\cdot,\theta_l \right)}^{\text{CL}}=\frac{w_{\mathcal{C}\left(\cdot,\theta_l \right)}^{\text{in}}+2\times \frac{1}{2}(k_l^w-1)-k_{l}^w}{1}+1=w_{\mathcal{C}\left(\cdot,\theta_l \right)}^{\text{in}},\\ 
		h_{\mathcal{C}\left(\cdot,\theta_l \right)}^{\text{CL}}\thinspace=\frac{h_{\mathcal{C}\left(\cdot,\theta_l \right)}^{\text{in}}\thinspace+2\times \frac{1}{2}(k_l^h-1)-k_{l}^h\thinspace}{1}+1=h_{\mathcal{C}\left(\cdot,\theta_l \right)}^{\text{in}}.
		\label{eq32}
	\end{cases}
\end{equation}
Based on Eq.(\ref{eq32}), the output width and height dimensions of the CL in $\mathcal{C}\left(\cdot,\theta_l \right) $ are $w_{\mathcal{C}\left(\cdot,\theta_l \right)}^{\text{in}}$  and $h_{\mathcal{C}\left(\cdot,\theta_l \right)}^{\text{in}}$, respectively. On the other hand, the BN and AL do not change the input dimension, i.e., the output width and height dimensions of the BN and AL in  $\mathcal{C}\left(\cdot,\theta_l \right) $ are also  $w_{\mathcal{C}\left(\cdot,\theta_l \right)}^{\text{in}}$  and $h_{\mathcal{C}\left(\cdot,\theta_l \right)}^{\text{in}}$, respectively. Consequently, the output width and height dimensions of $\mathcal{C}\left(\cdot,\theta_l \right) $ are  $w_{\mathcal{C}\left(\cdot,\theta_l \right)}^{\text{in}}$  and $h_{\mathcal{C}\left(\cdot,\theta_l \right)}^{\text{in}}$,  i.e., $w_{\mathcal{C}\left(\cdot,\theta_l \right)}^{\text{out}}=w_{\mathcal{C}\left(\cdot,\theta_l \right)}^{\text{in}}$ and  $h_{\mathcal{C}\left(\cdot,\theta_l \right)}^{\text{out}}=h_{\mathcal{C}\left(\cdot,\theta_l \right)}^{\text{in}}$, respectively. Besides, note that the convolution operation guarantees that the architectural parameters are positive integers.  That is, both $p_l^w$,  $p_l^h$ and $k_l^w$, $k_l^h$ are guaranteed to be positive integers during the convolution operation. $\hfill\blacksquare$

\emph{Proposition 2: When $s_l^w>1$, $s_l^h>1$, if $p_l^w=\frac{1}{2}(w_{\mathcal{C}\left(\cdot,\theta_l \right)}^{\text{in}}s_l^w-w_{\mathcal{C}\left(\cdot,\theta_l \right)}^{\text{in}}-s_l^w+k_l^w)$ and $p_l^h=\frac{1}{2}(h_{\mathcal{C}\left(\cdot,\theta_l \right)}^{\text{in}}s_l^h-h_{\mathcal{C}\left(\cdot,\theta_l \right)}^{\text{in}}-s_l^h+k_l^h)$, as well as $p_l^w$,  $p_l^h$, $s_l^w$, $s_l^h$, $k_l^w$, $k_l^h$ are positive integers, then $w_{\mathcal{C}\left(\cdot,\theta_l \right)}^{\text{out}}=w_{\mathcal{C}\left(\cdot,\theta_l \right)}^{\text{in}}$ and  $h_{\mathcal{C}\left(\cdot,\theta_l \right)}^{\text{out}}=h_{\mathcal{C}\left(\cdot,\theta_l \right)}^{\text{in}}$.}

\emph{Proof:} For the CL in $\mathcal{C}\left(\cdot,\theta_l \right) $, where $s_l^w>1$, $s_l^h>1$, $p_l^w=\frac{1}{2}(w_{\mathcal{C}\left(\cdot,\theta_l \right)}^{\text{in}}s_l^w-w_{\mathcal{C}\left(\cdot,\theta_l \right)}^{\text{in}}-s_l^w+k_l^w)$ and $p_l^h=\frac{1}{2}(h_{\mathcal{C}\left(\cdot,\theta_l \right)}^{\text{in}}s_l^h-h_{\mathcal{C}\left(\cdot,\theta_l \right)}^{\text{in}}-s_l^h+k_l^h)$, its output width and height dimensions $w_{\mathcal{C}\left(\cdot,\theta_l \right)}^{\text{CL}} \times h_{\mathcal{C}\left(\cdot,\theta_l \right)}^{\text{CL}} $ are denoted as
\begin{equation}
	\begin{cases}
		w_{\mathcal{C}\left(\cdot,\theta_l \right)}^{\text{CL}}=& \frac{w_{\mathcal{C}\left(\cdot,\theta_l \right)}^{\text{in}}+2\times \frac{1}{2}(w_{\mathcal{C}\left(\cdot,\theta_l \right)}^{\text{in}}s_l^w-w_{\mathcal{C}\left(\cdot,\theta_l \right)}^{\text{in}}-s_l^w+k_l^w)-k_{l}^w}{s_l^w}\\&+1
		=w_{\mathcal{C}\left(\cdot,\theta_l \right)}^{\text{in}},\\
		h_{\mathcal{C}\left(\cdot,\theta_l \right)}^{\text{CL}}=& \frac{h_{\mathcal{C}\left(\cdot,\theta_l \right)}^{\text{in}}+\thinspace2\times \frac{1}{2}(h_{\mathcal{C}\left(\cdot,\theta_l \right)}^{\text{in}}s_l^h-\thinspace h_{\mathcal{C}\left(\cdot,\theta_l \right)}^{\text{in}}-\thinspace s_l^h+\thinspace k_l^h)-\thinspace k_{l}^h}{s_l^h}\\&+1
		=h_{\mathcal{C}\left(\cdot,\theta_l \right)}^{\text{in}}.
		\label{eq33}
	\end{cases}
\end{equation}
Similarly, the output width and height dimensions of $\mathcal{C}\left(\cdot,\theta_l \right) $ are  $w_{\mathcal{C}\left(\cdot,\theta_l \right)}^{\text{in}}$  and $h_{\mathcal{C}\left(\cdot,\theta_l \right)}^{\text{in}}$,  i.e., $w_{\mathcal{C}\left(\cdot,\theta_l \right)}^{\text{out}}=w_{\mathcal{C}\left(\cdot,\theta_l \right)}^{\text{in}}$ and  $h_{\mathcal{C}\left(\cdot,\theta_l \right)}^{\text{out}}=h_{\mathcal{C}\left(\cdot,\theta_l \right)}^{\text{in}}$. Besides, it is also necessary to ensure that $p_l^w$,  $p_l^h$, $s_l^w$, $s_l^h$, and $k_l^w$, $k_l^h$ are positive integers during the convolution operation. $\hfill\blacksquare$

Subsequently, we incorporate \emph{Propositions 1} and \emph{2} to derive some architectural conditions that satisfy the beamforming properties in \emph{Remark 1}. Concretely, when $\hat{\mathbf{H}}_{\text{mod}}^{\text{3D}}\in \mathbb{R}^{Q \times I \times M}$ is inputted into $\mathcal{R}\left(\cdot,\theta \right) $,  the architectural parameters of the CL in the first $\mathcal{C}\left(\cdot,\theta_1 \right) $ are available in two cases. In the first case, when $s_1^w=1$ and $s_1^h=1$, $p_1^w$ and $p_1^h$ can be set to $\frac{1}{2}(k_1^w-1)$ and $\frac{1}{2}(k_1^h-1)$,  where both $p_1^w$,  $p_1^h$ and $k_1^w$, $k_1^h$ are positive integers. Based on \emph{Proposition 1}, the width and height dimensions of $\mathbf{V}_{\mathcal{C}\left(\cdot,\theta_1 \right)}$ are $Q\times I$.  Another case,  when $s_1^w>1$ and $s_1^h>1$, $p_1^w$ and $p_1^h$ can be set to $\frac{1}{2}(Qs_1^w-Q-s_1^w+k_1^w)$ and $\frac{1}{2}(Is_1^h-I-s_1^h+k_1^h)$, in which $p_1^w$,  $p_1^h$, $s_1^w$, $s_1^h$, and $k_1^w$, $k_1^h$ are positive integers. Based on \emph{Proposition 2}, the width and height dimensions of $\mathbf{V}_{\mathcal{C}\left(\cdot,\theta_1 \right)}$ are $Q\times I$. Similarly, as long as the architectural parameters $k_l^w$, $k_l^h$,  $p_l^w$, $p_l^h$, $s_l^w$, $s_l^h$, $l=1,\cdots, L$ in each $\mathcal{C}\left(\cdot,\theta_l \right) $ satisfy \emph{Proposition 1}  or \emph{Proposition 2}, the width and height dimensions of $\mathbf{V}_{\mathcal{C}\left(\cdot,\theta_L \right)}$ are $Q\times I$.
In addition, let $c_l$ denote the number of convolution kernels for the CL in $\mathcal{C}\left(\cdot,\theta_l \right) $. As long as the number of convolution kernels $c_L$ for  $\mathcal{C}\left(\cdot,\theta_L \right) $ is equal to $2M$, $\mathbf{V}_{\mathcal{C}\left(\cdot,\theta_L \right)}$ is a 3D real tensor of dimension $Q \times I \times 2M$. Besides, the dimension of the output  $\mathbf{V}_{\text{IM}}$ of the dentity  mapping is also $Q \times I \times 2M$, since the dentity mapping adjusts its architecture parameters to ensure that the dimension of $\mathbf{V}_{\text{IM}}$ is equal to that of $\mathbf{V}_{\mathcal{C}\left(\cdot,\theta_L \right)}$. Consequently, based on Eq.(\ref{eq21}),  $ \mathbf{V}_{\mathcal{R}\left(\cdot,\theta \right)}$  is a 3D real tensor of dimension $Q \times I \times 2M$. To sum up, the architectural conditions that satisfy the beamforming properties in \emph{Remark 1} are summarized in \emph{Remark 2}.

\emph{Remark 2: When $\hat{\mathbf{H}}_{\text{mod}}^{\text{3D}}\in \mathbb{R}^{Q \times I \times M}$ is fed into $\mathcal{R}\left(\cdot,\theta \right) $, $ \mathbf{V}_{\mathcal{R}\left(\cdot,\theta \right)}$  is a 3D real tensor of dimension $Q \times I \times 2M$ as long as the following two conditions are satisfied.
	\begin{itemize}
		\item[1)] The architectural parameters $k_l^w$, $k_l^h$,  $p_l^w$, $p_l^h$, $s_l^w$, $s_l^h$, $l=1,\cdots, L$ in each $\mathcal{C}\left(\cdot,\theta_l \right) $ satisfy Proposition 1 or Proposition 2.
		\item[2)] The number of convolutional kernels $c_L$ for  $\mathcal{C}\left(\cdot,\theta_L \right) $ is equal to $2M$.
	\end{itemize}}

In summary, based on \emph{Proposition 1}, \emph{Proposition 2} and \emph{Remark 2}, as long as the two conditions in \emph{Remark 2}  are satisfied, the output $ \mathbf{V}_{\mathcal{R}\left(\cdot,\theta \right)}$ of $\mathcal{R}\left(\cdot,\theta \right) $ is a 3D real tensor of dimension $Q \times I \times 2M$ by taking $\hat{\mathbf{H}}_{\text{mod}}^{\text{3D}}\in \mathbb{R}^{Q \times I \times M}$ as the input. This satisfies the beamforming properties in \emph{Remark 1}.

\subsection{Adaptive AP Clustering $\mathcal{A}\left(\cdot,\theta \right) $}
$\mathcal{R}\left(\cdot,\theta \right) $ achieves the mapping from $\hat{\mathbf{H}}_{\text{mod}}^{\text{3D}}\in \mathbb{R}^{Q \times I \times M}$ to  $\mathbf{V}_{\mathcal{R}\left(\cdot,\theta \right)} \in \mathbb{R}^{Q \times I \times 2M}$  as long as the two conditions in \emph{Remark 2}  are satisfied. In the following, on the basis of satisfying the two conditions in \emph{Remark 2}, $\mathcal{A}\left(\cdot,\theta \right) $ implements that  $\mathbf{V}_{\mathcal{R}\left(\cdot,\theta \right)} \in \mathbb{R}^{Q \times I \times 2M}$  contains more zero-blocks for AP clustering. To achieve this, one of the most intuitive ways is to feed the elements of $\mathbf{V}_{\mathcal{R}\left(\cdot,\theta \right)} \in \mathbb{R}^{Q \times I \times 2M}$ with a threshold function. That is,  when the  element of $\mathbf{V}_{\mathcal{R}\left(\cdot,\theta \right)} \in \mathbb{R}^{Q \times I \times 2M}$ is less than the threshold value of the threshold function, this element is set to 0, otherwise 1. Despite the simplicity of this approach, this suffers from two major problems.

\begin{itemize}
	\item[1)] The threshold value of the threshold function is usually set manually and empirically, which cannot change with the input 3D real CSI $\hat{\mathbf{H}}_{\text{mod}}^{\text{3D}}\in \mathbb{R}^{Q \times I \times M}$. However, the results of AP clustering vary with the input 3D real CSI $\hat{\mathbf{H}}_{\text{mod}}^{\text{3D}}\in \mathbb{R}^{Q \times I \times M}$, in which the threshold value of the threshold function in turn determines the results of AP clustering. Thus, for  AP clustering, the threshold value of the threshold function should vary with the input 3D real CSI $\hat{\mathbf{H}}_{\text{mod}}^{\text{3D}}\in \mathbb{R}^{Q \times I \times M}$.
	\item[2)] The threshold function is non-differentiable, which cannot be optimized along with $\mathcal{R}\left(\cdot,\theta \right) $ during the training period. Nevertheless, the objective of this paper is  robust joint AP clustering and beamforming design, which requires optimizing beamforming and  AP clustering simultaneously. Therefore, the threshold function should be able to be optimized along with $\mathcal{R}\left(\cdot,\theta \right) $ during the training period.
\end{itemize}

First, we address the first problem, i.e., making that the threshold value varies with the input 3D real CSI $\hat{\mathbf{H}}_{\text{mod}}^{\text{3D}}\in \mathbb{R}^{Q \times I \times M}$. To be specific, cell-free systems have the unrealistic drawback, i.e., long-range APs serving users consume precious power and bandwidth resources, while contributing little useful power due to high path losses\cite{bib4}. In other words,
for a user in cell-free systems, the CSI modulus
for the longer-range APs will usually be smaller than those of the shorter-range APs due to the larger path losses of the
longer-range APs. Accordingly, to reduce the above-mentioned unfavourable problem, when dealing with the CSI modulus corresponding to the longer-range APs, their corresponding threshold values can be set to smaller values for easier implementation of AP clustering, and vice versa. Consequently, $\mathcal{A}\left(\cdot,\theta \right) $ uses the spatial attention\cite{bib8} to realize that the threshold value changes with the 3D real CSI $\hat{\mathbf{H}}_{\text{mod}}^{\text{3D}}\in \mathbb{R}^{Q \times I \times M}$. It includes the pooling layer, $1\times1$ CL and AL. Formally, this is denoted as
\begin{equation}
\begin{aligned}
\mathbf{T}&=\text{AL}\left(\text{CL}_{1\times1}\left(\text{POOL}\left(\hat{\mathbf{H}}_{\text{mod}}^{\text{3D}}\right),\theta_\text{sa}\right)\right)\\&=\left(\begin{array}{ccc}
	t_{1}^{1} & \cdots & t_{I}^{1}\\
	\vdots & \ddots & \vdots\\
	t_{1}^{Q} & \cdots & t_{I}^{Q}
\end{array}\right),
\end{aligned}
\label{eq22}
\end{equation}
where $\text{POOL}\left(\cdot \right) $ denotes pooling according to the third dimension, i.e., the dimension of $\text{POOL}\left(\hat{\mathbf{H}}_{\text{mod}}^{\text{3D}}\right) $ is $Q \times I$. $\text{CL}_{1\times1}(\cdot,\cdot)$ denotes the $1\times1$ CL, in which $\theta_\text{sa}$ denotes the parameters of the spatial attention. $\mathbf{T} \in\mathbb{R}^{Q \times I} $  is a threshold value matrix for the input 3D real CSI $\hat{\mathbf{H}}_{\text{mod}}^{\text{3D}}\in \mathbb{R}^{Q \times I \times M}$, in which  $t_i^q$ is the threshold value of the $q^{th}$ AP to the $i^{th}$ user.

\emph{Proposition 3: For the $i^{th}$ user, if the $q^{th}$ AP is a longer-range AP and the $p^{th}$ AP is a shorter-range AP, then  $t_i^q<t_i^p$.}

\emph{Proof:} Please see Appendix A for the detailed proof. $\hfill\blacksquare$

Based on \emph{Proposition 3}, for those long-range APs occupying precious power and bandwidth resources while contributing little useful power to the user, Eq.(\ref{eq22}) enables adaptive setting smaller threshold values to make the AP clustering easier, thereby effectively reducing the unfavourable fact mentioned above. On the other hand, it is obvious that the threshold value  $\mathbf{T} \in\mathbb{R}^{Q \times I} $ in Eq.(\ref{eq22}) varies with the input 3D real CSI $\hat{\mathbf{H}}_{\text{mod}}^{\text{3D}}\in \mathbb{R}^{Q \times I \times M}$, where between each AP and each user is adaptively designed an AP clustering threshold.  In summary, Eq.(\ref{eq22}) effectively solves the first problem mentioned above.

In what follows, we address the second problem, i.e., making the threshold function differentiable to optimize along with $\mathcal{R}\left(\cdot,\theta \right) $ during the training period. Specifically,$\mathcal{A}\left(\cdot,\theta \right) $ proposes a differentiable threshold function, which is defined as
\begin{equation}
	\text{DT}_i^q=\frac{1}{1+e^{-k\left(pv_i^q-t_i^q \right) }},\forall i\in\mathcal{I},\forall q\in\mathcal{Q},
	\label{eq23}
\end{equation}
where $k$ denotes an amplification parameter. $pv_i^q=\text{POOL}\left(\mathbf{V}_{\mathcal{R}\left(\cdot,\theta \right)}[q,i,:]\right)$  denotes the value for the beamforming $\mathbf{V}_{\mathcal{R}\left(\cdot,\theta \right)}[q,i,:]$ of the $q^{th}$ AP to the $i^{th}$ user pooled by the third dimension. The schematic diagrams of the differentiable threshold function $\text{DT}_i^q$ at different $k$ and $t_i^q$ are shown in Figs.2 and 3, respectively.

As shown in Fig.2, when the amplification parameter $k$ is gradually increased, the differentiable threshold function $\text{DT}_i^q$ gradually approaches the ideal threshold function, where the amplification parameter $k$ is set 50 empirically. As shown in Fig.3, if $pv_i^q$  is less than the threshold $t_i^q$, then the value of the differentiable threshold function $\text{DT}_i^q$ is 0, otherwise 1. Combining Figs.2 and 3, the differentiable threshold function $\text{DT}_i^q$ is extremely approximated to the ideal threshold function and is differentiable, which is optimized along with  $\mathcal{R}\left(\cdot,\theta \right) $ during the training period. That is, the second difficulty mentioned above is effectively solved. In conclusion, $\mathcal{A}\left(\cdot,\theta \right) $ realizes  AP clustering, where the results of AP clustering are defined as
\begin{equation}
	\begin{aligned}
		\mathbf{C}_{\mathcal{A}\left(\cdot,\theta \right)}=\left(\begin{array}{ccc}
			\text{DT}_1^1 & \cdots & \text{DT}_I^1\\
			\vdots & \ddots & \vdots\\
			\text{DT}_1^Q & \cdots & \text{DT}_{I}^{Q}
		\end{array}\right).
	\end{aligned}
	\label{eq24}
\end{equation}

\begin{figure}[t]
	\centering
	\includegraphics[scale=0.55]{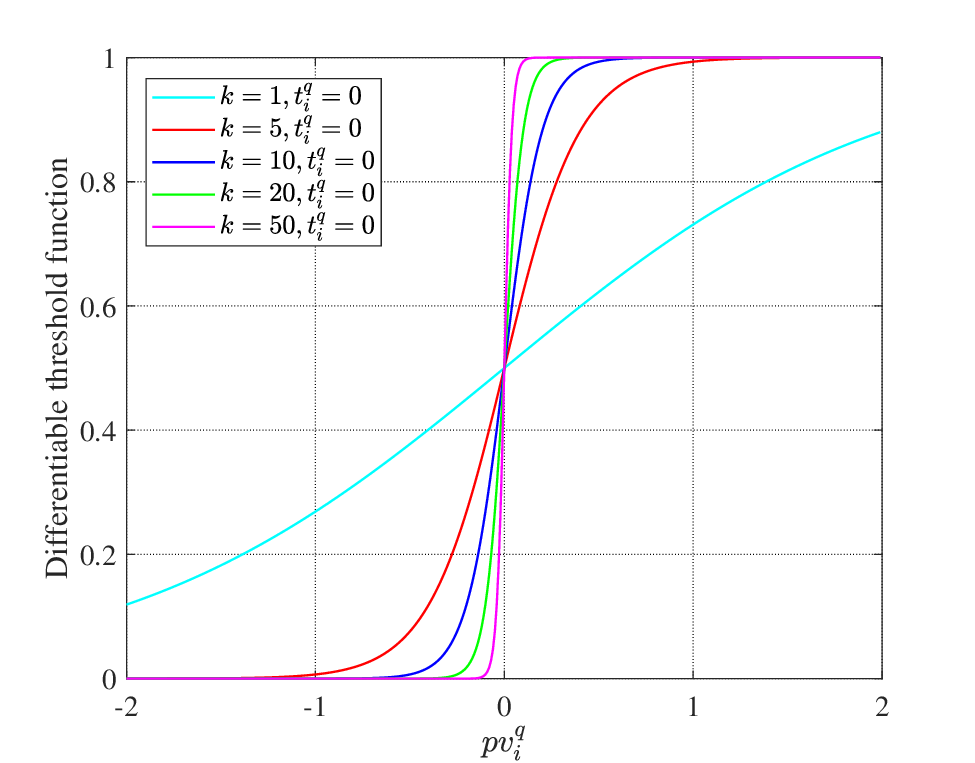}
	\caption{Differentiable threshold function (\ref{eq23}) at different  $ k $.} 
\end{figure}

Note that $\text{C2}$ in the optimization problem (\ref{eq19}) requires the beamforming vector to be 0 for those APs that are clustered as 0. To this end, the Hadamard product between $\mathbf{V}_{\mathcal{R}\left(\cdot,\theta \right)}$ and $ \mathbf{C}_{\mathcal{A}\left(\cdot,\theta \right)}$ is performed to obtain a sparse beamforming $\mathbf{V}_\text{spa}$, which is denoted as
\begin{equation}
\mathbf{V}_\text{spa}=\mathbf{V}_{\mathcal{R}\left(\cdot,\theta \right)} \otimes \mathbf{C}_{\mathcal{A}\left(\cdot,\theta \right)},
	\label{eq25}
\end{equation}
where $ \otimes $ denotes the Hadamard product of 2D matrix and 3D tensor. For example, for a 2D matrix $ \mathbf{A} \in \mathbb{R}^{a \times b } $ and a 3D tensor $ \mathbf{B} \in \mathbb{R}^{a \times b \times c } $, $ \mathbf{A} \otimes \mathbf{B} $ is calculated as follows. $ \mathbf{A} \in \mathbb{R}^{a \times b } $  is first copied $ c $ times to become $ \mathbf{C} \in \mathbb{R}^{a \times b \times c} $, and then the Hadamard product is performed on $ \mathbf{C} \in \mathbb{R}^{a \times b \times c} $ and $ \mathbf{B} \in \mathbb{R}^{a \times b \times c } $. Clearly, $\text{C2}$ in the optimization problem (\ref{eq19}) is satisfied since the elements in $ \mathbf{C}_{\mathcal{A}\left(\cdot,\theta \right)}$ are either 0 or 1.

\subsection{Beamforming Conversion $\mathcal{V}\left( \cdot\right) $ and Power Constraint $\mathcal{P}\left( \cdot\right) $} 
$\mathbf{V}_\text{spa}$ is a 3D real beamforming tensor of dimension $Q\times I \times 2M$, which should be transformed into a 3D complex beamforming tensor. For this purpose, $\mathcal{V}\left( \cdot\right) $ transforms $\mathbf{V}_\text{spa}$ into a 3D complex beamforming tensor as follows,
\begin{equation}
	\mathbf{V}_\text{spa}^{\text{com}}=\mathbf{V}_\text{S}[:,:,0:M]+j\mathbf{V}_\text{S}[:,:,M:2M],
	\label{eq26}
\end{equation}
where $\mathbf{V}_\text{spa}^{\text{com}}$ is a 3D complex beamforming tensor of dimension $Q\times I \times M$. On the other hand, $\mathbf{V}_\text{spa}^{\text{com}}$ also needs to satisfy the power constraint $\text{C1}$ of the optimization problem (\ref{eq19}). Due to the fact that the power constraint is a convex constraint\cite{bib9}, it can be satisfied using a projection function. Consequently, following \cite{bib9}, $\mathcal{P}\left( \cdot\right) $ applies the following projection function to satisfy the power constraint, i.e., 
\begin{equation}
	\mathbf{v}_i^q= \begin{cases}\mathbf{v}_i^q & \text { if } \underset{i\in \mathcal{I}}{\sum}\left(\mathbf{v}_{i}^{q}\right)^{\textrm{H}}\mathbf{v}_{i}^{q}\leq \text{P}_{\textrm{max}},\\ \frac{\mathbf{v}_i^q}{\underset{i\in \mathcal{I}}{\sum}\left(\mathbf{v}_{i}^{q}\right)^{\textrm{H}}\mathbf{v}_{i}^{q}} {\text{P}_{\text{max}}} & \text { otherwise, }\end{cases}
	\label{eq27}
\end{equation}
where $\mathbf{v}_i^q=\mathbf{V}_\text{spa}^{\text{com}}[q,i,:]$. Finally, the output beamforming of $\mathcal{P}\left( \cdot\right) $ is denoted as  $\mathbf{V}_{\text{RJAPCBN}}$. It is obvious that $\mathbf{V}_{\text{RJAPCBN}}$ satisfies  both  $\textrm{C1}$ and  $\textrm{C2}$ for the optimization problem (\ref{eq19}).
\subsection{Updating Slack Variable Module $\mathcal{U}\left( \cdot\right) $} In addition to satisfying $\textrm{C1}$ and $\textrm{C2}$, it is also necessary to satisfy  $\overline{\textrm{C4}}$, $\overline{\textrm{C5}}$, $\textrm{C6}$, where an unsupervised loss function also needs to be designed to train the proposed RJAPCBN for the purpose of  robust joint AP clustering and beamforming design with imperfect CSI in cell-free  systems. For this reason, $\mathcal{U}\left( \cdot\right) $ is proposed to fulfill 
$\overline{\textrm{C4}}$, $\overline{\textrm{C5}}$, $\textrm{C6}$ and to realize the closed-loop unsupervised training of the proposed RJAPCBN.

Recall that the optimization problem (\ref{eq19}),  $\overline{\textrm{C4}}$, $\overline{\textrm{C5}}$ and $\textrm{C6}$ determining the slack variable $\boldsymbol{{\gamma}}$ are simple convex constraints, where the elements of  $\overline{\textrm{C4}}$ and $\overline{\textrm{C5}}$ related to CSI and beamforming can be computed by $\hat{\mathbf{H}}$ and  $\mathbf{V}_{\text{RJAPCBN}}$. It encourages that a simple convex optimization problem can be solved to obtain the slack variable $\boldsymbol{{\gamma}}$ by the known  $\hat{\mathbf{H}}$ and  $\mathbf{V}_{\text{RJAPCBN}}$, which is denoted as
\begin{equation}
	\begin{split}
		\underset{\alpha_i,\beta_i,\gamma_i,\delta_i,\mu_i}{\textrm{max}}&\,\, \underset{i\in\mathcal{I}}{\sum}\log_2 \left(1+\gamma_{i}\right)
		\\\textrm{s.t.}&\quad
		\overline{\textrm{C4}},\overline{\textrm{C5}},\textrm{C6}.
	\end{split}
	\label{eq28}
\end{equation}
Note that $\underset{q\in \mathcal{Q}}{\sum}\left\Vert \mathbf{v}_{i}^{q}\right\Vert_1$ in the optimization problem (\ref{eq19}) is a constant without affecting the solution of the slack variables when knowing $ \mathbf{V}_{\text{RJAPCBN}}$, it is straightforward to remove in the optimization problem (\ref{eq28}) for simplicity. It is clear that the optimization problem (\ref{eq28}) is a simple convex optimization problem, which can be solved simply using the CVX toolbox in the MATLAB or the CVXPY toolbox in the python to obtain the optimal the slack variables $\boldsymbol{{\alpha}}^*$, $\boldsymbol{{\beta}}^*$, $\boldsymbol{{\delta}}^*$, $\boldsymbol{{\mu}}^*$ and $\boldsymbol{{\gamma}}^*$. Accordingly, with the slack variable $\boldsymbol{{\gamma}}^*$, the  unsupervised loss function of the proposed RJAPCBN can be defined as 
\begin{equation}
\mathcal{L}=-\underset{i\in\mathcal{I}}{\sum}\big(\log_2 \left(1+\gamma_{i}^*\right)-\lambda\underset{q\in \mathcal{Q}}{\sum}\left\Vert \mathbf{v}_{i}^{q}\right\Vert_1\big),
\label{eq29}
\end{equation}
where $\gamma_i^*$  is obtained by solving the optimization problem (\ref{eq28}). By minimizing $\mathcal{L}$ to unsupervised train the proposed RJAPCBN, the optimization problem (\ref{eq19}) is solved efficiently. In other words, with the above closed-loop unsupervised training,  the proposed  RJAPCBN realizes robust joint AP clustering and beamforming design with  imperfect CSI in cell-free  systems.

\begin{figure}[t]
	\centering
	\includegraphics[scale=0.55]{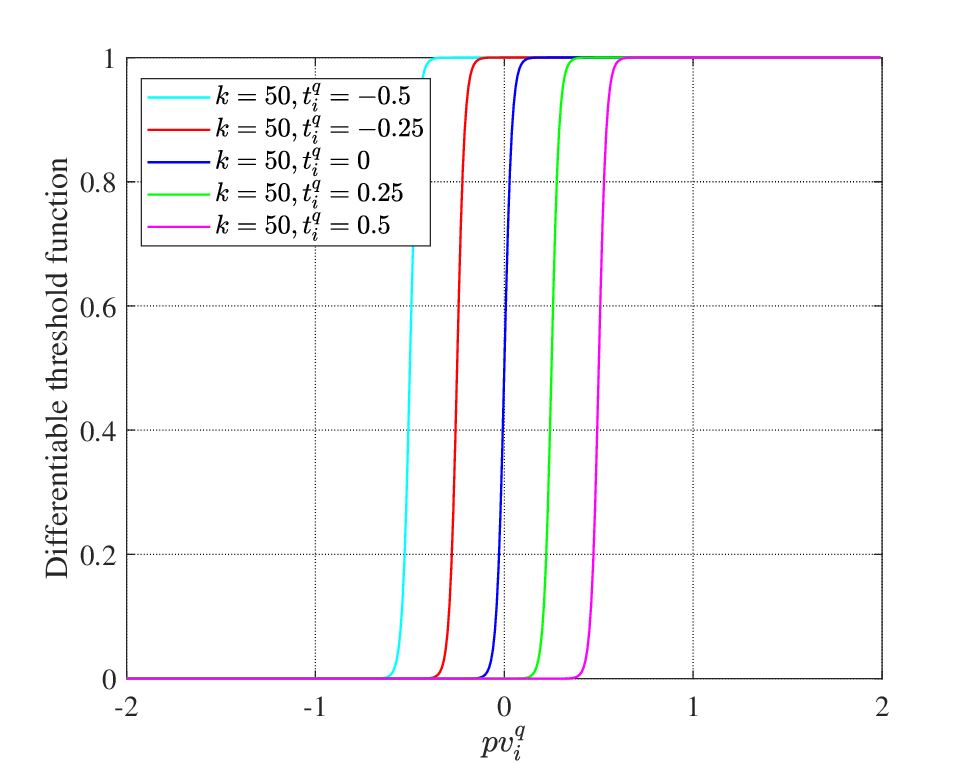}
	\caption{Differentiable threshold function (\ref{eq23}) at different  $ t_i^q $.} 
\end{figure}

\section{Experimental Results}
In this section, we validate the effectiveness of the proposed RJAPCBN in terms of parameter settings, the worst-case sum rate, the average number of serving APs per user and the computational complexity. The geographic location channel model \cite{bib20}, \cite{bib1} that is commonly exploited for beamforming design is selected, where the large-scale fading is modelled as $\left(200 / d_{i}^{q}\right)^{3} L_{i}^{q}$. Here, $d_{i}^{q} $ denotes the distance between the $q^{th}$ AP and the $i^{th}$ user, and $10 \log 10\left(L_{i}^{q}\right) \sim \mathcal{N}(0,64)$ denotes the shadowing effect. For ease of presentation, following \cite{bib12}, $\eta_{i}=\left\|\Delta\mathbf{h}_{i}\right\|_{2}/\left\|\mathbf{h}_{i}\right\|_{2},\forall i\in\mathcal{I}$ is defined as the error levels of imperfect CSI. In subsequent experiments, unless otherwise stated, the number of AP and users  is set to 16, where the number of antennas and the maximum power for each AP are set to 4 and 1, respectively. Besides, the performance of robust beamforming design  is measured by the commonly used worst-case sum rate. The performance of robust AP clustering can be measured by the average number of serving APs per user, which is defined as
\begin{equation}
	Q_{\text{ave}}=Q\left( 1-\frac{V_{\text{zero}}}{QIM} \right), 
	\label{eq30}
\end{equation}
where $V_{\text{zero}}$ denotes the number of zeros in
$\mathbf{V}_{\text{RJAPCBN}}$.  As $V_{\text{zero}}$ is larger, $Q_{\text{ave}} $ is smaller, i.e., the AP set of serving users is smaller, and vice versa. 

As benchmarks, the following schemes are compared:

$\bullet$ WMMSE with perfect CSI: The WMMSE \cite{bib23} achieves the stable solution of beamforming design in  perfect CSI by iteratively updating beamforming and a set of auxiliary variables. In addition, the WMMSE  is usually centralized to allow all APs to serve all users in cell-free  systems, which can be regarded as an upper bound of AP clustering. In summary, due to the excellent performance of the WMMSE with  perfect CSI, it can be viewed as an upper bound for robust joint AP clustering and beamforming design with imperfect CSI.

$\bullet$ WMMSE with imperfect CSI: The WMMSE \cite{bib23} directly treats imperfect CSI as perfect CSI  to highlight  the potential performance degradation caused by imperfect CSI.

$\bullet$ S-WMMSE with imperfect CSI: The S-WMMSE  \cite{bib20}  is a traditional optimization method for solving joint AP clustering and beamforming design under  perfect CSI, which  is also applied to imperfect CSI scenarios for highlighting the performance degradation that may result from imperfect CSI.

$\bullet$ CNNs with imperfect CSI:  \cite{bib22} applies the CNNs to implement AP clustering and beamforming design with perfect CSI individually, i.e., 
AP clustering is determined and beamforming is
then designed from the clustered APs. Similarly, the CNNs in \cite{bib22} is utilized to  imperfect CSI scenarios.

$\bullet$ JcbNet with imperfect CSI: The JcbNet \cite{bib1} is a deep learning method for joint AP clustering and beamforming design under  perfect CSI. Likewise, the JcbNet is applied to imperfect CSI scenarios.

\begin{table*}[t]
	\caption{The performance of the proposed RJAPCBN  under different architecture parameters.}
	\renewcommand{\arraystretch}{1.5}
	\centering
	\begin{tabular}{p{70pt}<{\centering}p{70pt}<{\centering}p{30pt}<{\centering}p{200pt}<{\centering}}
		\midrule[1pt]
		Convolution kernel&	Worst-case sum rate &	$Q_{\text{ave}} $&	Number of multiplications \\		
		\midrule[1pt]
		$7\times 7$& 264& 9.92& $Q^2I^2C+QIMC+QI+49QIMC+196QIC^2$\\	
		$7\times 5$& 256& 10.07& $Q^2I^2C+QIMC+QI+35QIMC+140QIC^2$\\
		$5\times 5$& 248& 10.24& $Q^2I^2C+QIMC+QI+25QIMC+100QIC^2$\\
		$5\times 3$& 235& 10.33& $Q^2I^2C+QIMC+QI+15QIMC+60QIC^2$\\
		$3\times 3$& 221& 10.54&$Q^2I^2C+QIMC+QI+9QIMC+36QIC^2$\\
		\midrule[1pt]
	\end{tabular}
\end{table*}

\subsection{Parameter Settings of RJAPCBN}
%$QIMC+QI+Q^2I^2C+QIMCK_1K_2+(L-1)QIC^2K_1K_2$

PyTorch is applied to implement the proposed RJAPCBN, where the Adam optimizer is selected.  The number of layers $L$ of $\mathcal{R}\left(\cdot,\theta \right) $ in the proposed RJAPCBN is set to 5. The learning rate and  batch size are set to 64 and 0.1, respectively. In the unsupervised training, 10000 channels are generated to train
the proposed RJAPCBN. In the model testing, 6400 channels are inputted into the trained RJAPCBN to output 3D complex beamforming. 

\begin{figure}[t]
	\centering
	\includegraphics[scale=0.55]{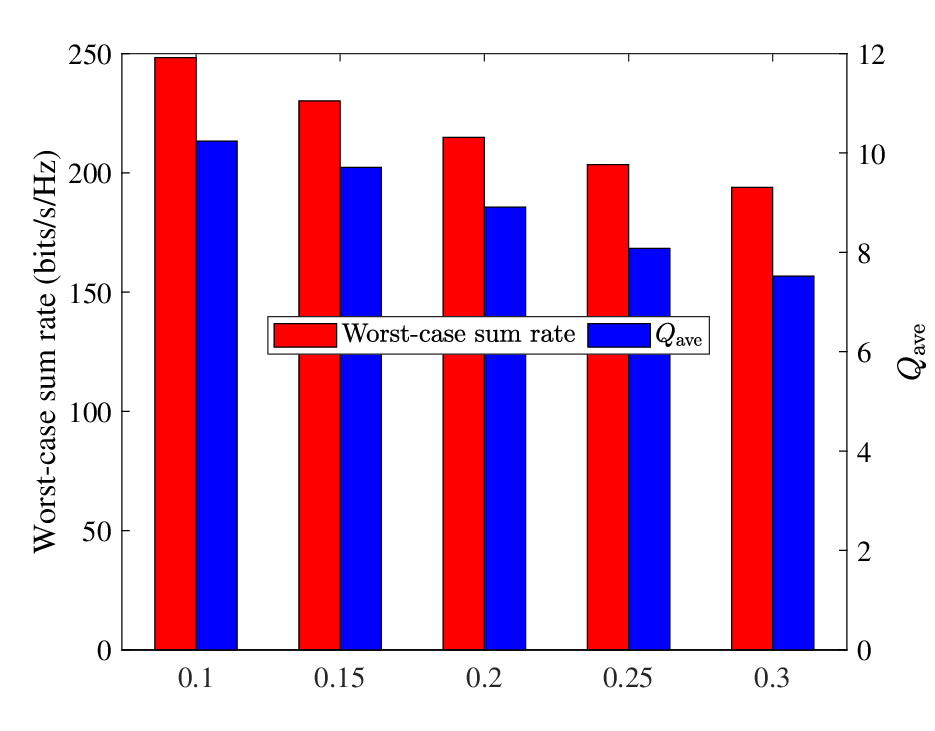}
	\caption{Worst-case sum rate and  $Q_{\text{ave}} $ at different  $\lambda$.}
	\label{fig4}
\end{figure}

From Eq. (\ref{eq19}), the hyperparameter $\lambda$ balances AP clustering and beamforming design, where the worst-case sum rate and  $Q_{\text{ave}} $  at different $\lambda$ are shown in Fig.\ref{fig4}. As $\lambda$ is larger, the worst-case sum rate and $Q_{\text{ave}} $ are smaller, and vice versa. For this reason,  $\lambda$ is selected to be 0.1, because the goal of this paper is to reduce $Q_{\text{ave}} $ as much as possible with minimal worst-case sum rate reduction. However, other scenarios allow flexibility in setting $\lambda$ according to different objectives.

According to \emph{Remark 2} in Section \uppercase\expandafter{\romannumeral3}.B, the proposed RJAPCBN could set different architectural parameters to realize robust joint AP clustering and beamfroming design with imperfect CSI in cell-free  systems, where the worst-case sum rate, $Q_{\text{ave}} $ and the number of multiplications for different architectural parameters are shown in TABLE  \uppercase\expandafter{\romannumeral1}. When the size of the convolution kernel decreases, the worst-case sum rate decreases and $Q_{\text{ave}} $ increases, while the computational complexity decreases, and vice versa. The reason is as follows: wireless communication channels often exhibit a block-sparse structure, i.e., a channel matrix that exhibits non-zero and zero values clustering structure\cite{bib26},\cite{bib27}. When the size of convolution kernel is reduced, the receptive field of convolution operation is reduced and easily dropped to a cluster of zero value, which results in the output of the convolution operation being close to zero. That is, less useful information is obtained, thus reducing the worst-case sum rate and increasing $Q_{\text{ave}} $, and vice verse.  Accordingly, the size of the convolution kernel of  the proposed RJAPCBN  is chosen $5\times 5$ in this paper, which is a balance between the computational complexity and the worst-case sum rate with $Q_{\text{ave}} $.

\begin{figure}[t]
	\centering
	\includegraphics[scale=0.55]{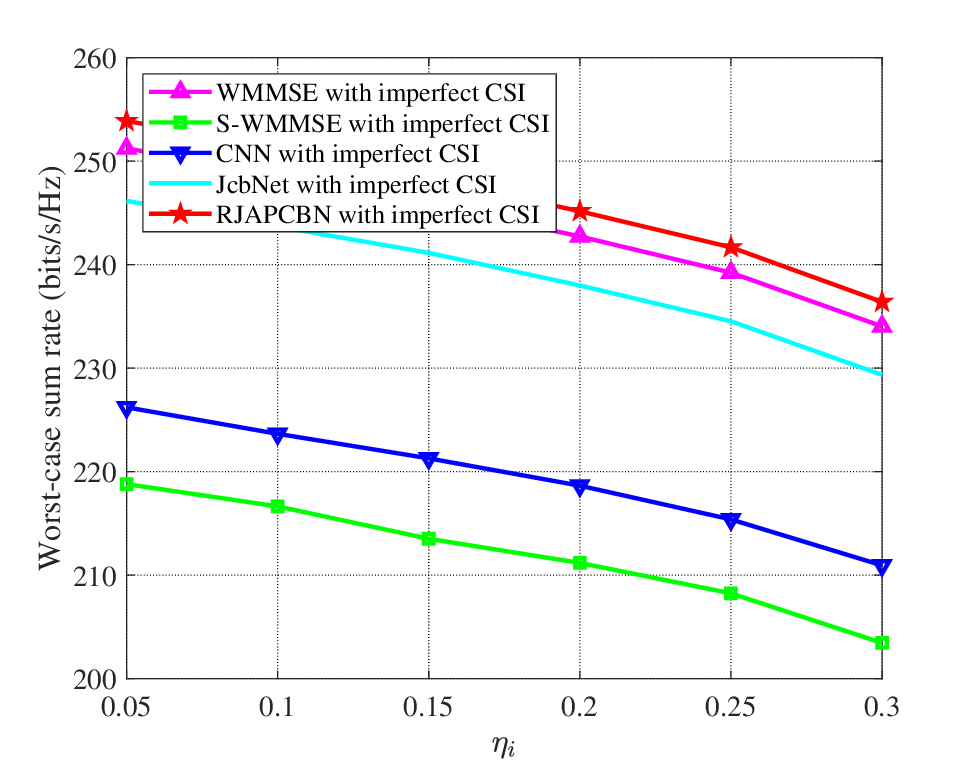}
	\caption{Worst-case sum rate  at different imperfect CSI error levels.}
	\label{fig11}
\end{figure} 

\begin{figure}[t]
	\centering
	\includegraphics[scale=0.55]{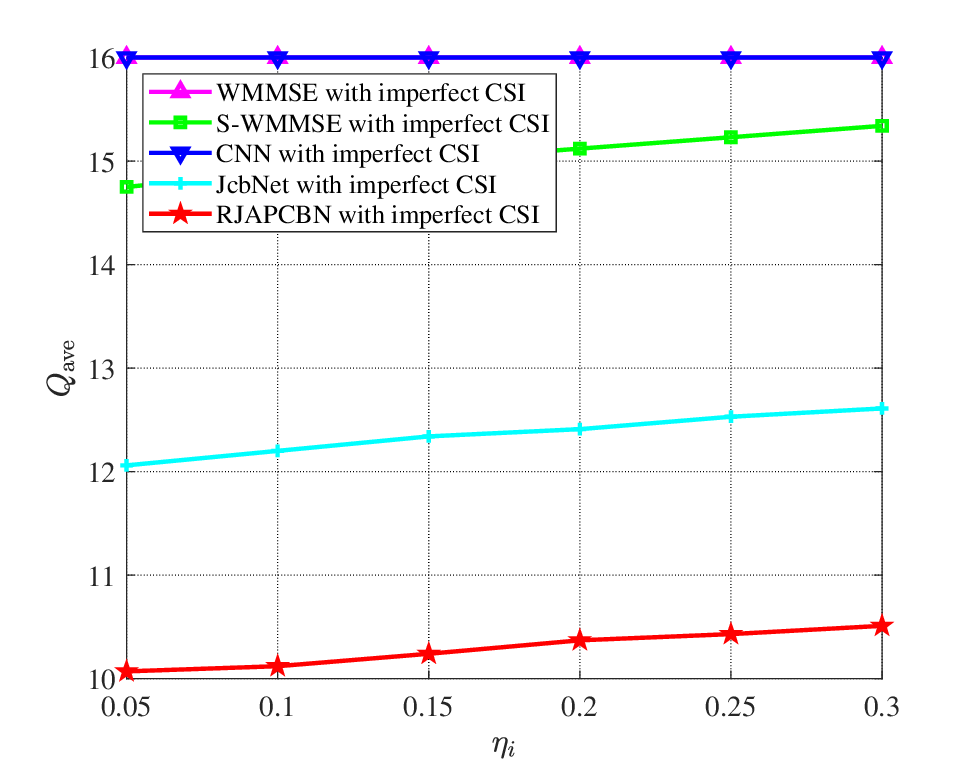}
	\caption{$Q_{\text{ave}}$ at different imperfect CSI error levels.}
	\label{fig12}
\end{figure}

\begin{figure}[t]
 	\centering
 	\includegraphics[scale=0.55]{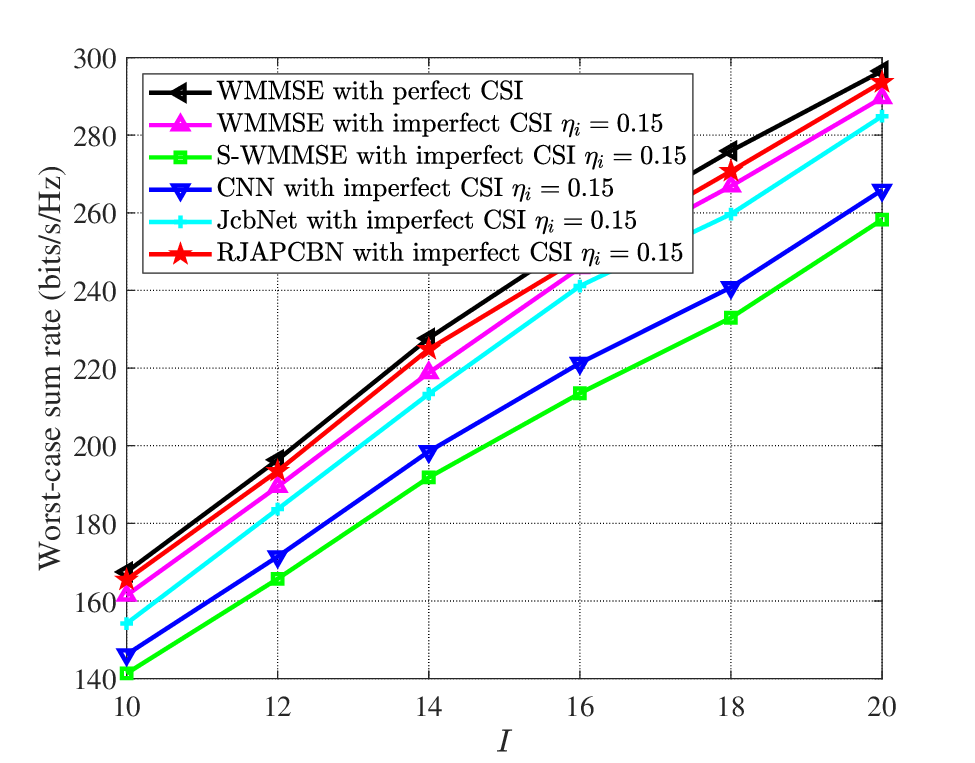}
 	\caption{Worst-case sum rate  at different number of users.}
 	\label{fig13}
\end{figure} 
 \begin{figure}[t]
 	\centering
 	\includegraphics[scale=0.55]{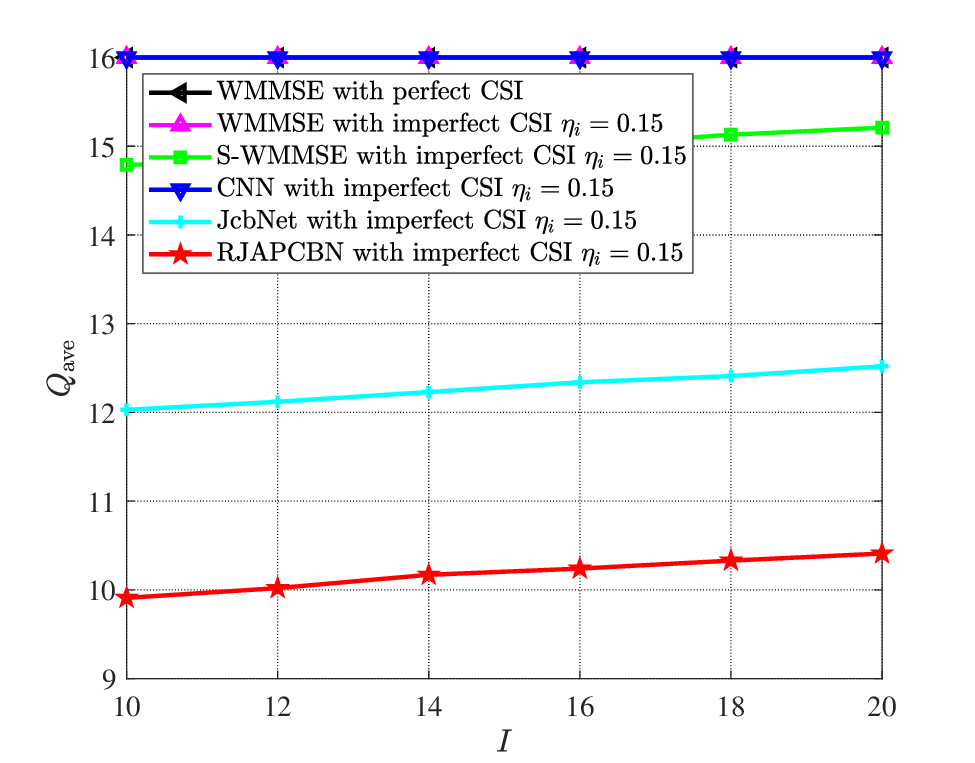}
 	\caption{$ Q_{\text{ave}} $ at  different number of users.}
 \end{figure}
 \begin{figure}[t]
 	\centering
 	\includegraphics[scale=0.55]{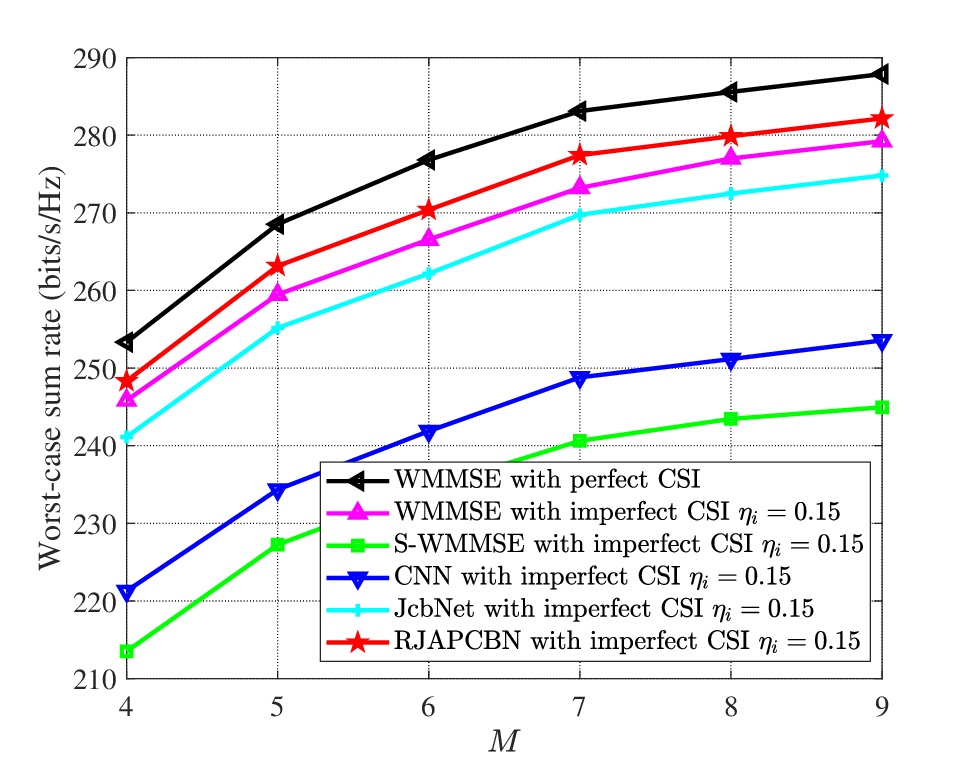}
 	\caption{Worst-case sum rate  at different no. of AP antennas.}
 \end{figure}
 
 \begin{figure}[t]
 	\centering
 	\includegraphics[scale=0.55]{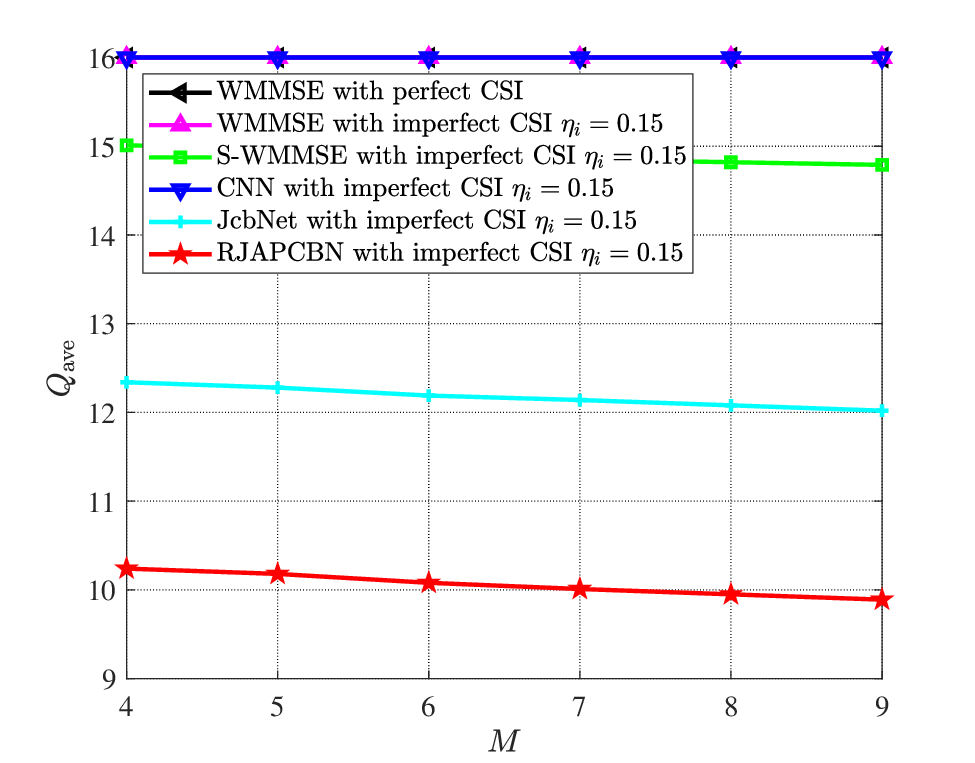}
 	\caption{$ Q_{\text{ave}} $ at different number of AP antennas.}
 	\label{fig14}
 \end{figure}
 
\begin{table*}[t]
	\caption{The computational complexity of several algorithms}
	\centering
	\renewcommand{\arraystretch}{1.5}
	\begin{tabular}{p{45pt}<{\centering}p{260pt}<{\centering}p{125pt}<{\centering}}
		\midrule[1pt]
		Algorithms&Number of multiplications&	Value \\		
		\midrule[1pt]
		WMMSE& $4L_{\textrm{ite}}(IQ^{3}M^{3}+I+I^{2}Q^{2}M^{2}+I^{2}+IQ^{2}M^{2}+IQM+4I^{2}QM+3IQM+IQM)$& $L_{\text {ite }}=15$ \\
		S-WMMSE& $4 L_{\text {ite }}(I+I^{2} +2 I Q^{2} M^{2}+2 I^{2} Q M +8 I Q M +I_{Q} Q(I(Q-1) M^{2}+I M+a I(\left(\log _{2} \epsilon\right)^{2}+1)(M^{3}+M^{2}+M)))$& $L_{\text {ite }}=15$, $ I_Q=10 $, $\epsilon=10^5$, $ a=0.9 $
		\\
		CNN& $QI(36MC+38KMC^{2}+LL_{l}+(2M+1)OO_{o})$& $ C=16 $, $ K=10 $, $L_l=QIM $, $ L=8 $, $ O=3 $, $ O_o=80 $\\
		JcbNet& $QIM+Q^2I^2M+QIMC+3QIMCk_l^wk_l^h+2QIMC(L-1)(2k_l^wk_l^h+1)+4QI_{\textrm{QoS}}MC$& $k_l^w=5$, $k_l^h=5$, $C=2M$, $L=5$\\	
		RJAPCBN& $Q^2I^2C+QIMC+QI+QIMCk_l^wk_l^h+(L-1)QIC^2k_l^wk_l^h$& $k_l^w=5$, $k_l^h=5$, $C=2M$, $L=5$\\		   
		\midrule[1pt]
	\end{tabular}
\end{table*}
 
\subsection{Performance of Worst-Case Sum Rate and Average Number of Serving APs Per User}

The worst-case sum rate and $Q_{\text{ave}} $ of these comparison algorithms at different imperfect CSI error levels $ \eta_{i}$ as well as number of users $I$ and AP antennas $M$ are shown Fig. 5-Fig.10, respectively. Under the same conditions, the worst-case sum rate of the proposed RJAPCBN with  imperfect CSI is higher than those of the  WMMSE, S-WMMSE, CNNs, JcbNet with imperfect CSI, which is approaching to the WMMSE with  perfect CSI. On the other hand, the $Q_{\text{ave}} $ of the proposed RJAPCBN with imperfect CSI is also lower than those of the S-WMMSE, CNNs, JcbNet with imperfect CSI, much lower than those of the WMMSE with  imperfect and perfect CSI. To summarize, compared to these algorithms, the proposed RJAPCBN with imperfect CSI achieves better worst-case sum-rate performance with a smaller  $Q_{\text{ave}} $. The reasons are as follows: The WMMSE with perfect CSI is a stable solution of beamforming design, thus its worst-case  sum rate is the highest. When faced with imperfect CSI  scenarios, the worst-case sum rate of the WMMSE degrades, because the WMMSE is designed without considering the robustness of imperfect CSI  scenarios. On the other hand, the WMMSE is all APs serving all users in cell-free systems, i.e., $Q_{\text{ave}} $ is the largest.  The S-WMMSE is also designed based on perfect CSI, where the worst-case sum rate decreases and $Q_{\text{ave}} $ increases when faced with imperfect CSI  scenarios. The CNNs are designed for AP clustering and beamforming design with perfect CSI individually, which is difficult to achieve the optimal solution. The JcbNet is joint AP clustering and beamforming design in perfect CSI scenarios, which also does not take into account the robustness of imperfect CSI scenarios. On the contrary, in addition to being designed for joint AP clustering and beamforming, the proposed RJAPCBN also takes into account the effect of imperfect CSI in the optimization problem. This effectively improves the robustness of imperfect CSI scenarios. Thus, the proposed RJAPCBN achieves better worst-case sum-rate performance with a smaller  $Q_{\text{ave}} $ with imperfect CSI.

\begin{figure}[t]
	\centering
	\includegraphics[scale=0.55]{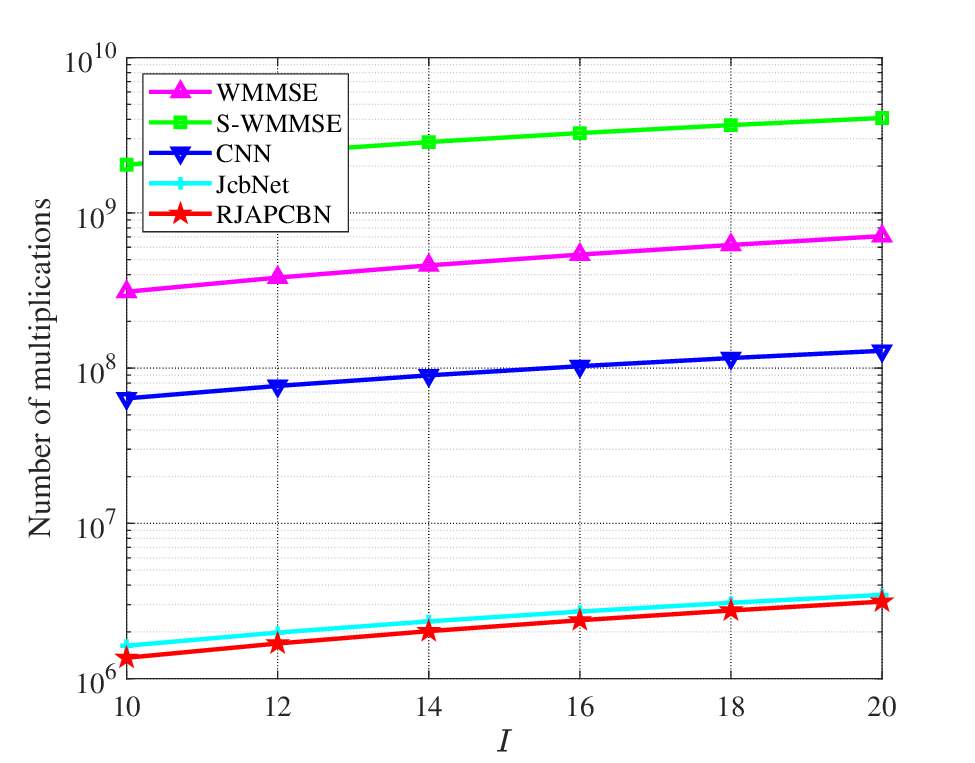}
	\caption{Computational complexity at different no. of users.}
	\label{fig17}
\end{figure}

\subsection{Computational Complexity}

Table \uppercase\expandafter{\romannumeral2} shows the  computational complexity  of several algorithms. To better compare the computational complexity, the number of multiplications of several algorithms under different number of users is shown in Fig.11. The number of multiplications for the S-WMMSE,  WMMSE, CNNs, JcbNet and  the proposed RJAPCBN is about $10^9$, $10^8$, $10^7\sim10^8$, $10^6$ and $10^6$, where the computational complexity of the proposed RJAPCBN is the lowest. The reasons are as follows: The S-WMMSE needs multiple matrix inversions and binary searches with high computational complexity, thus it has the highest computational complexity. The WMMSE also requires multiple matrix inversions, hence its computational complexity is higher. The CNNs reduces the computational complexity compared to the WMMSE and S-WMMSE. Nevertheless, the CNNs also contains the FC layers with a high number of neurons, which increases the computational complexity. Conversely, the JcbNet and the proposed RJAPCBN apply the  CL with parameter sharing mechanism  to  effectively reduce the computational complexity. Since the number of convolution units in each layer of  the proposed  RJAPCBN  is less than that of the JcbNet, the computational complexity of  the proposed  RJAPCBN  is smaller than that of the JcbNet. In summary, the proposed RJAPCBN  is a low-complexity robust
joint AP clustering and beamforming design method.

\section{Conclusion}
In this paper, a low-complexity unsupervised deep learning method RJAPCBN is proposed for roubst joint  AP clustering and beamforming design with imperfect CSI in cell-free  systems. The proposed RJAPCBN mainly includes the CSI conversion, residual network, adaptive AP clustering, beamforming conversion, power constraint and updating slack variable modules, which are combined for closed-loop unsupervised training to automatically find the optimal AP clustering and beamforming design with imperfect CSI in cell-free systems.  Numerical results demonstrated that the proposed   RJAPCBN achieves a higher worst-case sum rate under a smaller number of AP clustering with high computational efficiency.

\section*{Appendix A}
\section*{Proof of Proposition 3}
In cell-free systems, long-range APs serving users consume precious power and bandwidth resources, while contributing little useful power due to high path losses\cite{bib4}. In other words, for the $i^{th}$ user, if the $q^{th}$ AP is a longer-range AP and the $p^{th}$ AP is a shorter-range AP, then  $\text{POOL}\left(\mathbf{H}_{\text{mod}}^{\text{3D}}[q,i,:]\right)< \text{POOL}\left(\mathbf{H}_{\text{mod}}^{\text{3D}}[p,i,:]\right)$ due to the fact that the path loss of the $q^{th}$ AP is higher than that of the $p^{th}$ AP. On the other hand, the $1\times 1$ CL has the parameter-sharing mechanism, i.e., the parameters of $1\times 1$ CL are the same for $\text{POOL}\left(\mathbf{H}_{\text{mod}}^{\text{3D}}[q,i,:]\right) $ and $\text{POOL}\left(\mathbf{H}_{\text{mod}}^{\text{3D}}[q,i,:]\right) $. Consequently, based on Eq.(\ref{eq22}), it is convenient to obtain $t_i^q<t_i^p$. Thus,  we complete the proof of \emph{Proposition 3}.

% or




\ifCLASSOPTIONcaptionsoff
\newpage
\fi



%\bibliographystyle{IEEEtranTCOM}
%\bibliographystyle{IEEEtran}
\bibliographystyle{IEEEtran}
\bibliography{IEEEabrv,ref}

%\begin{biography}[{\includegraphics[width=1in,height=1.25in,clip,keepaspectratio]{mshell}}]{Michael Shell}


%\vfill


\end{document}